\documentclass[usenatbib]{mn2e}

\usepackage{graphicx}
\usepackage{amsmath}
\usepackage{lscape}
\usepackage{afterpage}
\usepackage{soul}
\usepackage{color}

\title[The ultraluminous state revisited]{The ultraluminous state revisited: fractional variability and spectral shape as diagnostics of super-Eddington accretion}

\author[A.\,D. Sutton et al.]{Andrew D. Sutton$^1$\thanks{Email: andrew.sutton@durham.ac.uk}, Timothy P. Roberts$^1$, Matthew J. Middleton$^{1, 2}$\\
 \\
$^1$Department of Physics, University of Durham, South Road, Durham, DH1 3LE, UK\\
$^2$Astronomical Institute Anton Pannekoek, Science Park 904, 1098 XH, Amsterdam, Netherlands
}

\pagerange{\pageref{firstpage}--\pageref{lastpage}} \pubyear{2011}

\def\Msun{\hbox{$\rm M_{\odot}$}}

\def\ergsec{{\rm ~erg~s^{-1}}}

\def\H0{{\rm ~km~s^{-1}~Mpc^{-1}}}

\def\la{\mathrel{\hbox{\rlap{\hbox{\lower4pt\hbox{$\sim$}}}{\raise2pt\hbox{$<$}}
}}}
\def\ga{\mathrel{\hbox{\rlap{\hbox{\lower4pt\hbox{$\sim$}}}{\raise2pt\hbox{$>$}}
}}}
\def\d25{$D_{25}$}

\begin{document}

\maketitle

\label{firstpage}

\begin{abstract}

Although we are nearing a consensus that most ultraluminous X-ray
sources (ULXs) below 10$^{41}$ erg s$^{-1}$ represent stellar-mass
black holes accreting in a super-Eddington `ultraluminous' accretion
state, little is yet established of the physics of this extreme 
accretion mode.  Here, we use a combined X-ray spectral and timing
analysis of an {\it XMM-Newton} sample of ULXs to investigate this new
accretion regime.  We start by suggesting an empirical classification
scheme that separates ULXs into three classes based on the spectral
morphologies observed by \cite{gladstone_etal_2009}: a singly-peaked
{\it broadened disc\/} class, and two-component {\it hard ultraluminous\/}
and {\it soft ultraluminous\/} regimes, with the spectra of the
latter two classes dominated by the harder and softer component
respectively.  We find that at the lowest luminosities ($L_{\rm X} < 3
\times 10^{39} \rm ~erg~s^{-1}$) the ULX population is dominated by
sources with broadened disc spectra, whilst ULXs with two component
spectra are seen almost exclusively at higher luminosities, suggestive
of a distinction between $\sim$ Eddington and super-Eddington
accretion modes.  We find high levels of fractional variability are
limited to ULXs with soft ultraluminous spectra, and a couple of
the broadened disc sources.  Furthermore, the variability in these
sources is strongest at high energies, suggesting it originates in the
harder of the two spectral components.  We argue that these properties
are consistent with current models of super-Eddington emission, where
a massive radiatively-driven wind forms a funnel-like geometry around
the central regions of the accretion flow.  As the wind provides the
soft spectral component this suggests that inclination is the key
determinant in the observed two-component X-ray spectra, which is very
strongly supported by the variability results if this originates due
to clumpy material at the edge of the wind intermittently obscuring
our line-of-sight to the spectrally hard central regions of the ULX.
The pattern of spectral variability with luminosity in two ULXs that
straddle the hard/soft ultraluminous regime boundary is
consistent with the wind increasing at higher accretion rates, and
thus narrowing the opening angle of the funnel.  Hence, this work
suggests that most ULXs can be explained as stellar-mass black holes
accreting at and above the Eddington limit, with their observed
characteristics dominated by two variables: accretion rate and
inclination. 

\end{abstract}

\begin{keywords}
accretion, accretion discs -- black hole physics -- X rays: binaries -- 
X rays: galaxies
\end{keywords}

\section{Introduction}\label{intro}

\begin{table*}
\caption{The ULX sample}
\centering
\begin{tabular}{cccccc}
\hline
Source name & Reference$^a$ & 2XMM ID & ${N_{\rm H}}^b$ & Distance$^c$ & Reference$^d$ \\
\hline
NGC 55 ULX    & 1, 2, 3          & 2XMM J001528.8$-$391318 & 1.73 & 2.11 & i \\
M31 ULX1      & 4, 5, 6          & CXOM31 J004253.1$+$411422$^e$ & 6.68 & 0.79 & i \\
NGC 253 XMM2  & 3, 7             & 2XMM J004722.6$-$252050 & 1.38 & 3.68 & i \\
NGC 253 ULX2  & 2, 7, 3          & 2XMM J004732.9$-$251749 & 1.38 & 3.68 & i \\
M33 X-8       & 1                & 2XMM J013350.8$+$303937 & 5.69 & 0.92 & i \\
NGC 1313 X-1  & 1, 2, 3, 7, 8    & 2XMM J031819.8$-$662911 & 3.95 & 4.39 & ii \\
NGC 1313 X-2  & 1, 2, 8          & 2XMM J031822.1$-$663603 & 3.95 & 4.39 & ii \\
IC 342 X-1    & 1, 3, 7, 8       & 2XMM J034555.6$+$680455 & 28.75 & 3.50 & iii \\
IC 342 X-2    & 3                & 2XMM J034606.5$+$680705 & 28.75 & 3.50 & iii \\
NGC 2403 X-1  & 1, 2, 3, 7, 8, 9 & 2XMM J073625.5$+$653540 & 4.17 & 3.50 & i\\
Ho II X-1     & 1, 2, 3, 7, 8, 9 & 2XMM J081928.9$+$704219 & 3.41 & 3.42 & ii \\
M81 X-6       & 1, 3, 8, 9       & 2XMM J095532.9$+$690034 & 4.22 & 4.27 & i \\
Ho IX X-1     & 1, 2, 8          & 2XMM J095753.2$+$690348 & 4.26 & 3.61 & iv \\
NGC 4190 ULX1 & 8                & CXO  J121345.2$+$363754$^e$ & 1.62 & 3.47 & v \\
NGC 4559 ULX2 & 1, 2, 3, 7, 8, 9 & 2XMM J123551.7$+$275604 & 0.80 & 6.98 & i \\
NGC 4736 ULX1 & 3                & 2XMM J125048.6$+$410743 & 1.44 & 4.66 & ii \\
NGC 5204 X-1  & 1, 2, 3, 7, 8, 9 & 2XMM J132938.6$+$582506 & 1.38 & 4.65 & vi \\
NGC 5408 X-1  & 1, 2, 3 , 7      & 2XMM J140319.6$-$412258 & 5.93 & 4.80 & vii \\
NGC 5907 ULX  & 3, 7             & 2XMM J151558.6$+$561810 & 1.38 & 13.4 & i \\
NGC 6946 X-1  & 3, 8, 9          & 2XMM J203500.0$+$601130 & 20.05 & 6.80 & viii \\
\hline
\end{tabular}
\begin{minipage}{\linewidth}
$^a$The sample sources are selected from: 
1 - \cite{gladstone_etal_2009}; 
2 - \cite{heil_etal_2009}; 
3 - Middleton et al. (in prep.); 
4 - \cite{henze_etal_2009}; 
5 - \cite{kaur_etal_2012};
6 - \cite{middleton_etal_2012};
7 - \cite{walton_etal_2011b};
8 - \cite{liu_and_bregman_2005}; 
9 - \cite{swartz_etal_2004}.
$^b$Galactic column density in the direction of the source ($\times 10^{20}~{\rm cm}^{-2}$), taken from \cite{dickey_and_lockman_1990}.
$^c$Distance to the ULX host galaxy in units of Mpc.
$^d$Distances to the ULX host galaxies are taken from: 
i - \cite{EDD}; 
ii - \cite{EDD2}; 
iii - \cite{herrmann_etal_2008}; 
iv - \cite{dalcanton_etal_2009}; 
v - \cite{tikhonov_and_karachentsev_1998}; 
vi - \cite{karachentsev_etal_2003}; 
vii - \cite{karachentsev_etal_2002}; 
viii - \cite{karachentsev_etal_2000}.
$^e$All {\it XMM-Newton} detections of M31 ULX1 and NGC 4190 ULX1 were obtained too late for these two sources to be included in the 2XMM-DR3 catalogue, so do not have a 2XMM source designation, in these cases {\it Chandra} source IDs are shown instead.
\end{minipage}
\label{sample}
\end{table*}

Ultraluminous X-ray sources (ULXs) are point sources of X-rays, with luminosities at or in excess of the Eddington limit for a typical Galactic black hole of mass $\sim 10 \Msun$ ($\ga 10^{39} \ergsec$).  They are located outside of the nucleus of their host galaxies, so cannot be powered by accretion onto super-massive black holes (SMBHs; $> 10^{5} \Msun$).  Whilst accretion onto a population of intermediate mass black holes (IMBHs; \citealt{colbert_and_mushotzky_1999}; $10^2 \Msun \la M_{\rm BH} \la 10^4 \Msun$) remains a possible explanation for a subset of some of the brightest ULXs (e.g. \citealt{farrell_etal_2009}; \citealt{sutton_etal_2012}), it is unlikely to be the case in the majority of the population.  The evidence for this is varied; notably if all ULXs contained IMBH primaries, it would imply an unfeasibly high IMBH formation rate in starforming galaxies \citep{king_2004}; and a population of IMBHs alone is inconsistent with the break at $\sim 2 \times 10^{40} \rm ~erg~s^{-1}$ in the luminosity function of point X-ray sources in starforming galaxies (e.g. \citealt{swartz_etal_2004}; \citealt{mineo_etal_2012}).  Indeed, this is strongly supported by the detection of both a soft excess and high energy curvature in a number of high quality {\it XMM-Newton} ULX observations (e.g. \citealt{roberts_etal_2005}; \citealt{stobbart_etal_2006}; \citealt{goncalves_and_soria_2006}; \citealt{gladstone_etal_2009}; \citealt{miyawaki_etal_2009}), which are inconsistent with these sources being in known sub-Eddington accretion states.  Instead, the majority of ULXs are likely powered by accretion onto fairly typical stellar remnant black holes.  In this case, many may be classic stellar mass black holes (sMBHs; $M_{\rm BH} < 20 \Msun$; \citealt{feng_and_soria_2011}), which would need to be accreting at super-Eddington rates to produce the observed luminosities.  However, this requirement could be relieved in some ULXs if they instead contain slightly larger massive stellar remnant black holes (MsBHs; $20 \Msun < M_{\rm BH} < 100 \Msun$; \citealt{feng_and_soria_2011}), which may form in regions of low metallicity (\citealt{zampieri_and_roberts_2009}; \citealt{mapelli_etal_2010}; \citealt{belczynski_etal_2010}).

\cite{stobbart_etal_2006} demonstrated that some of the highest quality {\it XMM-Newton} ULX spectra were well represented by a model consisting of two thermal (or thermal-like) components, different to the spectra shown by typical sub-Eddington black holes, suggestive of a new, super-Eddington `ultraluminous' accretion state \citep{roberts_2007}.  \cite{gladstone_etal_2009} then investigated the physics of the highest quality {\it XMM-Newton} ULX spectra in more detail, and in doing so identified three spectral types of ULXs, which they speculated could be placed into a sequence with increasing accretion rate.  In their interpretation, at around the Eddington limit ULXs appear with broad disc-like spectra; as the accretion rate becomes more super-Eddington a two-component spectrum emerges, initially peaking in the higher energy component; and, at the highest super-Eddington accretion rates the balance of the two components shifts, and the ULX appears instead with its peak at the softer end of the 0.3-10\,keV spectrum.  Here, we refer to these three ultraluminous spectral regimes as the {\it ``broadened disc'', ``hard ultraluminous''\/} and {\it ``soft ultraluminous''} states respectively (see Fig.~\ref{eg_spec} for an illustration of these spectra).  Initially \cite{gladstone_etal_2009} interpreted the hard and soft spectral components as a cool, optically thick corona around the inner disc, and the unobscured outer disc emission respectively, however more recent work has refined this.  The hard spectral component may indeed originate in a Comptonising corona around the inner regions of the accretion disc; similarly, the consistently super-Eddington Galactic black hole binary GRS 1915+105 \citep*{done_etal_2004} also requires cool Comptonisation (\citealt{zdziarski_etal_2001}; \citealt{middleton_etal_2009}; \citealt*{ueda_etal_2009}).  Alternatively, the hard emission could instead be coming directly from the hot inner disc (\citeauthor{middleton_etal_2011a} 2011a), with its spectral shape plausibly due to a large colour correction \citep{kajava_etal_2012}.  The soft component likely originates in the photosphere at the base of a massive radiatively driven wind, as is expected to occur in black holes that are close to or exceeding Eddington-limited accretion \citep{poutanen_etal_2007}; this is consistent with the apparent cooling of the soft component as its luminosity increases ($L \propto T^{-3.5}$; \citealt{kajava_and_poutanen_2009})\footnote{Other scenarios exist that can also produce the observed luminosity -- temperature relation, e.g. \cite{soria_2007}.}.  So, given this revision in our understanding of the nature of the two components, it is pertinent to ask whether the different regimes really are a function of accretion rate alone, or whether they might instead depend on other characteristics of the ULX.

One way to make progress in this regard is to ask the question: how do the observed short-term variability properties of ULXs relate to the spectral regimes seen in the ultraluminous state?  The extended thermal emission from the radiatively driven wind should be intrinsically stable over short time periods, given it originates in an optically-thick region.  However, the smaller scale of the inner emission region means that it could vary, although this is not the case in many ULXs which instead are observed to have suppressed variability \citep{heil_etal_2009}, possibly due to the stabilising effect of advection on the disc structure (\citealt{abramowicz_etal_1988}).  Despite this, a few ULXs have been reported as displaying short time-scale variability, including reported quasi-periodic variability in M82 X-1 \citep{strohmayer_and_mushotzky_2003}, NGC 5408 X-1 (\citealt{strohmayer_etal_2007}; \citealt{strohmayer_and_mushotzky_2009}; \citeauthor{middleton_etal_2011a} 2011a) and NGC 6946 X-1 \citep{rao_etal_2010}.  While the timescales of these QPOs have been cited as evidence of the presence of IMBHs (\citealt{strohmayer_and_mushotzky_2003}; \citealt{strohmayer_etal_2007}; \citealt{strohmayer_and_mushotzky_2009}; \citealt{rao_etal_2010}), \citeauthor{middleton_etal_2011a} (2011a) have suggested a way in which such variability could be produced in the ultraluminous state.  If the material in the wind is clumpy in nature \citep{takeuchi_etal_2013}, then if the line of sight to the central regions intersects the edge of the wind the clumps will imprint stochastic variability onto the hard component by intermittently obscuring it.  As the soft emission originates in the wind itself, it is not adversely affected by this mechanism (\citeauthor{middleton_etal_2011a} 2011a;  Middleton et al. in prep.).

However, this model of super-Eddington emission is not universally accepted, and other ULX models may explain the current data.  Reflection models \citep{caballero_garcia_and_fabian_2010} give one alternative interpretation of the energy spectrum from ULXs.  These require a radiation source with a power-law-like continuum to be produced above a high spin black hole; the proponents of such models have suggested that power could be magnetically extracted from the disc and released in the emission region, thus avoiding the Eddington limit \citep{caballero_garcia_and_fabian_2010}.  The observed ULX spectral features, including the soft excess and high energy break, are then produced by relativistically-blurred reflection from the disc.  Investigating the variability characteristics of ULXs provides a good test of such models -- the detection of multiple spectral components, each with different variability properties, would be difficult to explain in terms of reflection.  A further proposed model of ULX emission is that of slim (super-critical) discs \citep{abramowicz_etal_1988}; these can reproduce the flat power-law spectra seen in many ULXs, and predict black hole masses in the massive stellar regime \citep{vierdayanti_etal_2006}.  However, it is not clear how variability could be produced in a slim disc alone, again providing a potential diagnostic test of such models; an additional physical issue is that slim disc spectral models neglect the wind that would be expected to arise as the accretion rate exceeds the Eddington limit (\citealt{poutanen_etal_2007}; \citealt{dotan_and_shaviv_2011}), although later slim disc simulations do self-consistently include winds (e.g. \citealt{ohsuga_and_mineshige_2011}).

Here we present results from a combined X-ray spectral and timing analysis using multiple {\it XMM-Newton} detections of a sample of 20 ULXs with moderate to high quality X-ray data.  We suggest a new empirical scheme to classify the ULXs by X-ray spectrum into three ultraluminous spectral types: broadened disc, hard ultraluminous and soft ultraluminous.  Then, we characterise the spectral and timing properties of ULXs in each class in order to further explore and constrain models of accretion in ULXs.

\section{Sample selection and data reduction}

\begin{table*}
\caption{Observation log}
\centering
\begin{tabular}{cccccccccc}
\hline
Obs. ID$^a$ & Date$^b$ & ${t_{\rm exp}}^c$ & Count rate$^d$ & $\theta^e$ & Obs. ID$^a$ & Date$^b$ & ${t_{\rm exp}}^c$ & Count rate$^d$ & $\theta^e$ \\
 & & (ks) & $({\rm ct~s^{-1}})$ & (arcmin) & & & (ks) & $({\rm ct~s^{-1}})$ & (arcmin) \\
\hline
\multicolumn{10}{c}{NGC 55 ULX} \\
0028740201     & 2001-11-14 & 30.4 & 2.0 & 4.40 &
0028740101     & 2001-11-15 & 21.4 & 1.1 & 11.25 \\
0655050101     & 2010-05-24 & 83.0 & 1.1 & 1.13 \\
\\
\multicolumn{10}{c}{M31 ULX1} \\
0600660201$^f$ & 2009-12-28 & 16.2 & 5.2 & 2.66 &
0600660301$^f$ & 2010-01-07 & 15.4 & 5.2 & 2.53 \\
0600660401     & 2010-01-15 &  6.8 & 5.1 & 2.47 &
0600660501     & 2010-01-25 & 11.4 & 4.1 & 2.38 \\
0600660601     & 2010-02-02 & 10.8 & 3.4 & 2.27 \\
\\
\multicolumn{10}{c}{NGC 253 XMM2} \\
0125960101$^g$ & 2000-06-03 & 34.2 & 0.1 & 5.37 &
0110900101     & 2000-12-13 &  7.4 & 0.1 & 4.55 \\
0152020101     & 2003-06-19 & 38.0 & 0.4 & 5.29 &
0304850901     & 2006-01-02 &  9.8 & 0.2 & 3.12 \\
0304851001     & 2006-01-06 &  9.8 & 0.2 & 3.15 &
0304851201     & 2006-01-09 & 18.0 & 0.2 & 3.17 \\
\\
\multicolumn{10}{c}{NGC 253 ULX2} \\
0125960101     & 2000-06-03 & 34.2 & 0.4 & 1.67 &
0110900101     & 2000-12-13 &  7.4 & 0.5 & 3.11 \\
0152020101     & 2003-06-19 & 38.0 & 0.4 & 1.61 &
0304850901     & 2006-01-02 &  9.8 & 0.4 & 0.80 \\
0304851001     & 2006-01-06 &  9.8 & 0.4 & 0.79 &
0304851201     & 2006-01-09 & 18.0 & 0.4 & 0.77 \\
0304851301$^g$ & 2006-01-11 &  6.4 & 0.2 & 0.80 \\
\\
\multicolumn{10}{c}{M33 X-8} \\
0102640401$^g$ & 2000-08-02 & 11.6 & 1.2 & 12.29 &
0102640101     & 2000-08-04 &  5.6 & 8.4 & 1.16 \\
0102640701$^g$ & 2001-07-05 & 10.0 & 1.3 & 12.92 &
0102641001$^g$ & 2001-07-08 &  6.0 & 1.6 & 10.50 \\
0102642001     & 2001-08-15 &  8.4 & 2.2 & 14.08 &
0102642101     & 2002-01-25 &  9.8 & 4.2 & 10.65 \\
0102642301     & 2002-01-27 & 10.0 & 4.5 & 8.67 &
0141980601     & 2003-01-23 & 10.0 & 2.6 & 14.05 \\
0141980801     & 2003-02-12 &  6.6 & 6.9 & 1.10 &
0141980101$^g$ & 2003-07-11 &  6.2 & 1.6 & 10.62 \\
0141980301$^h$ & 2003-07-25 &  6.4 & 2.9 & 8.56 &
0650510101     & 2010-07-09 & 60.6 & 3.0 & 9.53 \\
0650510201     & 2010-07-11 & 53.6 & 7.1 & 4.06 \\
\\
\multicolumn{10}{c}{NGC 1313 X-1} \\
0106860101     & 2000-10-17 & 10.4 & 1.1 & 1.45 &
0150280301$^h$ & 2003-12-21 &  6.8 & 0.9 & 7.87 \\
0150280601     & 2004-01-08 &  6.0 & 1.2 & 7.76 &
0205230301     & 2004-06-05 &  8.2 & 1.7 & 5.63 \\
0205230501$^g$ & 2004-11-23 & 15.4 & 0.4 & 7.83 &
0205230601     & 2005-02-07 &  4.4 & 0.9 & 7.38 \\
0405090101     & 2006-10-15 & 77.4 & 1.0 & 1.51 \\
\\
\multicolumn{10}{c}{NGC 1313 X-2} \\
0106860101     & 2000-10-17 & 10.4 & 0.4 & 5.44 &
0150280301     & 2003-12-21 &  6.8 & 1.4 & 1.02 \\
0150280601     & 2004-01-08 &  6.0 & 0.6 & 1.04 &
0205230301     & 2004-06-05 &  8.2 & 1.5 & 1.26 \\
0205230501     & 2004-11-23 & 12.6 & 0.5 & 0.99 &
0205230601     & 2005-02-07 &  4.4 & 1.5 & 1.10 \\
0301860101     & 2006-03-06 & 18.2 & 1.0 & 6.17 &
0405090101     & 2006-10-15 & 77.4 & 0.9 & 5.40 \\
\\
\multicolumn{10}{c}{IC 342 X-1} \\
0093640901     & 2001-02-11 &  5.6 & 0.5 & 5.07 &
0206890201     & 2004-08-17 & 18.0 & 0.6 & 4.26 \\
\\
\multicolumn{10}{c}{IC 342 X-2} \\
0093640901$^h$ & 2001-02-11 &  5.6 & 0.2 & 4.25 &
0206890201     & 2004-08-17 & 18.0 & 0.4 & 1.86 \\
\\
\multicolumn{10}{c}{NGC 2403 X-1} \\
0164560901     & 2004-09-12 & 56.0 & 0.2 & 5.36 \\
\\
\multicolumn{10}{c}{Ho II X-1} \\
0112520601     & 2002-04-10 &  5.0 & 4.4 & 1.12 &
0200470101     & 2004-04-15 & 21.0 & 4.5 & 1.13 \\
0561580401     & 2010-03-26 & 24.0 & 1.7 & 1.13 \\
\\
\multicolumn{10}{c}{M81 X-6} \\
0111800101$^i$ & 2001-04-22 & 77.2 & 0.6 & 3.35 &
0112521001     & 2002-04-10 &  7.0 & 0.5 & 11.93 \\
0112521101     & 2002-04-16 &  8.0 & 0.5 & 11.88 &
0200980101$^g$ & 2004-09-26 & 61.0 & 0.2 & 13.85 \\

\hline
\end{tabular}
\begin{minipage}{\linewidth}
$^a${\it XMM-Newton} observation identifiers.
$^b$Observation start date, in yyyy-mm-dd format.
$^c$The amount of simultaneous good time in 200s bins, in all of the EPIC detectors used in the analysis.
$^d$Combined {\it XMM-Newton} EPIC count rate of the ULX.
$^e$Angular separation between the on-axis position of the observation and the 2XMM source position.
$^f$No MOS2 detection was included in the analysis of this observation.
$^g$No PN detection was included in the analysis of this observation.
$^h$No MOS1 or MOS2 detections were included in the analysis of this observation.
$^i$No MOS1 detection was included in the analysis of this observation.
\end{minipage}
\label{obs_log}
\end{table*}

\begin{table*}
\begin{flushleft} 
{\bf Table \ref{obs_log}.} (continued)
\end{flushleft}
\centering
\begin{tabular}{cccccccccc}
\hline
Obs. ID$^a$ & Date$^b$ & ${t_{\rm exp}}^c$ & Count rate$^d$ & $\theta^e$ & Obs. ID$^a$ & Date$^b$ & ${t_{\rm exp}}^c$ & Count rate$^d$ & $\theta^e$ \\
 & & (ks) & $({\rm ct~s^{-1}})$ & (arcmin) & & & (ks) & $({\rm ct~s^{-1}})$ & (arcmin) \\
\hline

\multicolumn{10}{c}{Ho IX X-1} \\
0111800101$^g$ & 2001-04-22 & 77.2 & 0.7 & 12.534 &
0112521001     & 2002-04-10 &  7.0 & 2.9 & 1.11 \\
0112521101     & 2002-04-16 &  8.0 & 3.3 & 1.13 &
0200980101     & 2004-09-26 & 25.0 & 2.3 & 1.13 \\
\\
\multicolumn{10}{c}{NGC 4190 ULX1} \\
0654650201     & 2010-06-08 &  6.0 & 1.5 & 1.09 &
0654650301     & 2010-11-25 &  8.8 & 2.3 & 1.18 \\
\\
\multicolumn{10}{c}{NGC 4559 ULX2} \\
0152170501     & 2003-05-27 & 21.0 & 0.5 & 1.08 \\
\\
\multicolumn{10}{c}{NGC 4736 ULX1} \\
0404980101     & 2006-11-27 & 37.0 & 0.4 & 2.05 \\
\\
\multicolumn{10}{c}{NGC 5204 X-1} \\
0142770101     & 2003-01-06 & 12.0 & 0.8 & 1.13 &
0405690101     & 2006-11-15 &  7.6 & 1.7 & 1.10 \\
0405690201     & 2006-11-19 & 22.0 & 1.5 & 1.08 &
0405690501     & 2006-11-25 & 17.0 & 1.0 & 1.13 \\
\\
\multicolumn{10}{c}{NGC 5408 X-1} \\
0112290601     & 2001-08-08 &  4.8 & 1.7 & 1.27 &
0112290701$^g$ & 2001-08-24 &  5.6 & 0.7 & 1.23 \\
0302900101     & 2006-01-13 & 88.0 & 1.2 & 1.09 &
0500750101     & 2008-01-13 & 27.3 & 1.1 & 1.06 \\
0653380201     & 2010-07-17 & 64.2 & 1.5 & 1.12 & 
0653380301     & 2010-07-19 & 106.2 & 1.5 & 1.12 \\
0653380401     & 2011-01-26 & 83.4 & 1.4 & 1.11 &
0653380501     & 2011-01-28 & 86.2 & 1.4 & 1.07 \\
\\
\multicolumn{10}{c}{NGC 5907 ULX} \\
0145190201     & 2003-02-20 &  8.2 & 0.5 & 2.40 &
0145190101     & 2003-02-28 &  9.8 & 0.4 & 2.41 \\
0673920301     & 2012-02-09 & 13.0 & 0.2 & 1.12 \\
\\
\multicolumn{10}{c}{NGC 6946 X-1} \\
0200670101$^g$ & 2004-06-09 &  4.0 & 0.2 & 1.38 &
0200670301     & 2004-06-13 &  8.0 & 0.5 & 1.41 \\
0200670401$^g$ & 2004-06-25 &  6.0 & 0.2 & 1.37 &
0500730201     & 2007-11-02 & 28.4 & 0.4 & 3.54 \\
0500730101     & 2007-11-08 & 18.0 & 0.4 & 3.53 \\

\hline
\end{tabular}
\label{obslog}
\end{table*}

\subsection{Sample selection}\label{sample_selection}

The primary driver of our ULX sample selection was the requirement for the data to be of high enough quality to allow us to conduct a statistically significant short-term timing analysis; as such, we defined the criteria for selecting observations based on combined EPIC count rates and the available good time.  As a starting point, we used the ULXs previously identified as those with the highest quality {\it XMM-Newton} data, which were subject to an in-depth spectral analysis by \cite{gladstone_etal_2009}, and power spectral density analysis by \cite{heil_etal_2009}.  In addition to these, we also considered ULXs with fluxes in excess of $5 \times 10^{-13}~{\rm erg~cm^{-2}~s^{-1}}$ in the {\it ROSAT} HRI observed sample of \cite{liu_and_bregman_2005}, and the {\it Chandra} archival sample of \cite{swartz_etal_2004} as these are likely candidates for having reasonable quality {\it XMM-Newton} data, if they have been detected in previous observations.  To these we added M31 ULX1 (CXOM31 J004253.1+411422; \citealt{henze_etal_2009}; \citealt{kaur_etal_2012}; \citealt{middleton_etal_2012}), and all ULXs with a 0.3--10 keV count rate of $\ga 0.3~{\rm ct~s^{-1}}$ in at least one observation in a cross correlation of the 2XMM-DR3 \citep{watson_etal_2009} and the RC3 \citep{deVaucouleurs_etal_1991} catalogues (Middleton et al. in prep.; which is an updated version of \citealt{walton_etal_2011b}).

We then obtained and examined all of the archival {\it XMM-Newton} observations of these ULXs.  Firstly we extracted their combined EPIC count rates.  Motivated by the available ULX data, it was decided to extract fractional variability from light curves with 200~s temporal binning, as a compromise between having $\ga 20$ counts per bin and at least 20 temporal bins in a large sample of observations.  When extracting the fractional variability we use a combined light curve from all of the available EPIC detections of the source, so we define the available good time in an observation as the sum of the 200~s bins in which all of the available detectors were on and free from high levels of background flaring.  We then rejected all of the sources which did not meet the selection criteria of a minimum average 0.3--10 keV combined EPIC count rate of $0.3~{\rm ct~s^{-1}}$, in at least one {\it XMM-Newton} observation, and sufficient good time to extract light curves with at least twenty 200s bins (i.e. a total of 4~ks good time), to ensure Gaussian statistics.  The ULXs remaining in the sample are listed in Table \ref{sample}, along with the Galactic column densities and distances to the assumed host galaxy.  For the ULXs which met the sample criteria in at least one observation, we included all observations that had the same minimum good time, but met a less stringent count rate limit of at least $0.1~{\rm ct~s^{-1}}$, corresponding to an average of 20 counts per bin.  In addition to full-band (0.3--10 keV) timing analysis, we also examine the properties in two other bands (0.3-1 and 1-10 keV); for this we use sub-sets of the observations which meet the count rate limit of $\ge 0.1~{\rm ct~s^{-1}}$ in the appropriate energy band.  

\subsection{Data reduction}\label{DR}

All archival {\it XMM-Newton} EPIC detections of the ULXs shown in Table~\ref{sample} were downloaded from the NASA HEASARC archive\footnote{\tt http://heasarc.gsfc.nasa.gov/docs/archive.html}, and those which met the criteria described in section~\ref{sample_selection} were included in the subsequent work.  A proprietary observation of NGC 5907 ULX also met these criteria, so was included in the analysis \citep{sutton_etal_2013}.  We summarise all the data used in our analyses in Table~\ref{obslog}.

The data were reduced, and products extracted using standard tools in {\it XMM-Newton} {\sc sas}\footnote{\tt http://xmm.esac.esa.int/sas/} (version 10.0.0).  
Firstly, we extracted high energy (10--15 keV) full field light curves using {\sc evselect}, from which we filtered out periods subject to high background flaring, using {\sc tabgtigen} to generate good time interval (GTI) files.  When creating GTIs, we ensured that only temporally complete bins were used (i.e. no drop-outs were left in the data that could affect the subsequent timing analysis).  The exact filtering count rate varied between observations, to maximise the utilised data whilst avoiding flaring, but typical values were $\sim$ 1--1.5 and $\sim$ 0.6 $\rm ct~s^{-1}$ for the PN and MOS detectors respectively.  When later calculating fractional variability we required that all of the available detectors were simultaneously on, so defined a GTI using the detector with the latest start time and earliest stop time.  Generally this was the PN detector, except for observations with no PN detection of the ULX; here either MOS1 or MOS2 light curves were used to define the GTI.

Source spectra and light curves were mainly extracted from circular source regions with radii between 20--50 arcseconds, with apertures at the lower end of the range being used when other sources were in close proximity to the object of interest.  One exception, where neighbouring sources necessitated a very small source region radius of 12.5 arcseconds, was NGC 253 ULX2.  Also, in a few cases elliptical source regions were used, which were mainly necessary when the object neighboured a detector chip gap\footnote{The affected observations were: NGC 55 ULX - 0028740101 (PN); M33 X-8 - 0102642301 (PN, MOS1 \& MOS2) and 0650510101 (PN, MOS1 \& MOS2); NGC 1313 X-1 - 0150280601 (MOS1 \& MOS2); NGC 1313 X-2 - 0106860101 (MOS1 \& MOS2), 0301860101 (MOS1 \& MOS2) and 0405090101 (MOS1 \& MOS2); NGC 2403 X-1 - 0164560901 (PN); M81 X-6 - 0111800101 (PN \& MOS2); NGC 6946 X-1 - 0200670301 (PN), 0500730201 (PN) and 0500730101 (PN).}.  An elliptical region was also used for the MOS1 detection of M81 X-6 in observation 0112521001, where it was aligned with the direction of maximum point spread function as the source image was particularly elongated due to its large off-axis angle.  One further exception was observation 0600660201 of M31 ULX1, here we followed \cite{middleton_etal_2012}, and used an annular source region to account for pile-up in the PN detection.  Again following \cite{middleton_etal_2012}, we excluded the MOS2 data from the analysis of observations 0600660201 and 0600660301, as significant pile-up in these observations would have required the removal of large centroids.  Count rates were low enough that pile-up was not an issue in other observations.  Background spectra/light curves were extracted from circular regions; these were set equal in area to the source regions when extracting light curves, but larger regions were used when extracting spectra.  Background regions were located at a similar distance from the read out node as the source, on the same or a neighbouring chip for PN detections, or on the same chip as the source for MOS detections.

Final data products were extracted using the appropriate GTI file, standard event patterns ({\tt PATTERN} $\le 4$ for the PN detector, {\tt PATTERN} $\le 12$ for the MOS detectors) and filters ({\tt FLAG} $= 0$ for spectra; {\tt \#xmmea\_ep} or {\tt \#xmmea\_em} respectively for the PN and MOS light curves).  Spectra, and appropriate response matrices were extracted using {\sc xmmselect}, then grouped using the {\sc ftool}\footnote{\tt https://heasarc.gsfc.nasa.gov/ftools/} {\sc grppha}; light curves were obtained using {\sc evselect}.

\section{Analysis and Results}

\subsection{An empirical spectral classification scheme for ULXs}

\begin{figure}
\centering
\includegraphics[width=8.cm, angle=0]{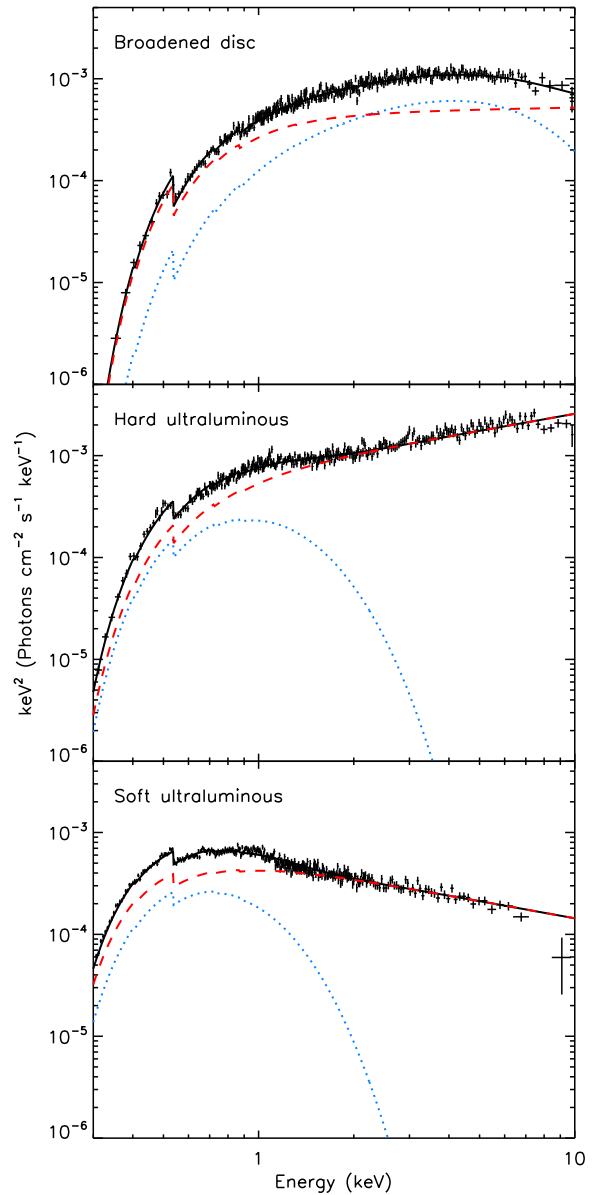}
\caption{Example spectra from observations of different ULXs exhibiting each of the three spectral types.  From top to bottom they are: broadened disc - observation 0405090101 of NGC 1313 X-2; hard ultraluminous - observation 0200980101 of Ho IX X-1 and soft ultraluminous - observation 0653380301 of NGC 5408 X-1.  For clarity, data are rebinned to $10 \sigma$ significance and only EPIC PN detections are shown.  The contributions from each of the components in the best fitting absorbed MCD (blue dotted line) plus power-law (red dashed line) model are shown.  The subtle spectral turnover indicative of ultraluminous state spectra is clearly seen in the data in the lower two panels, falling below the power-law tail whose slope is predominantly set by the data between $\sim 2$ and 6 keV.}
\label{eg_spec}
\end{figure}

\begin{figure}
\centering
\includegraphics[width=8.cm, angle=0]{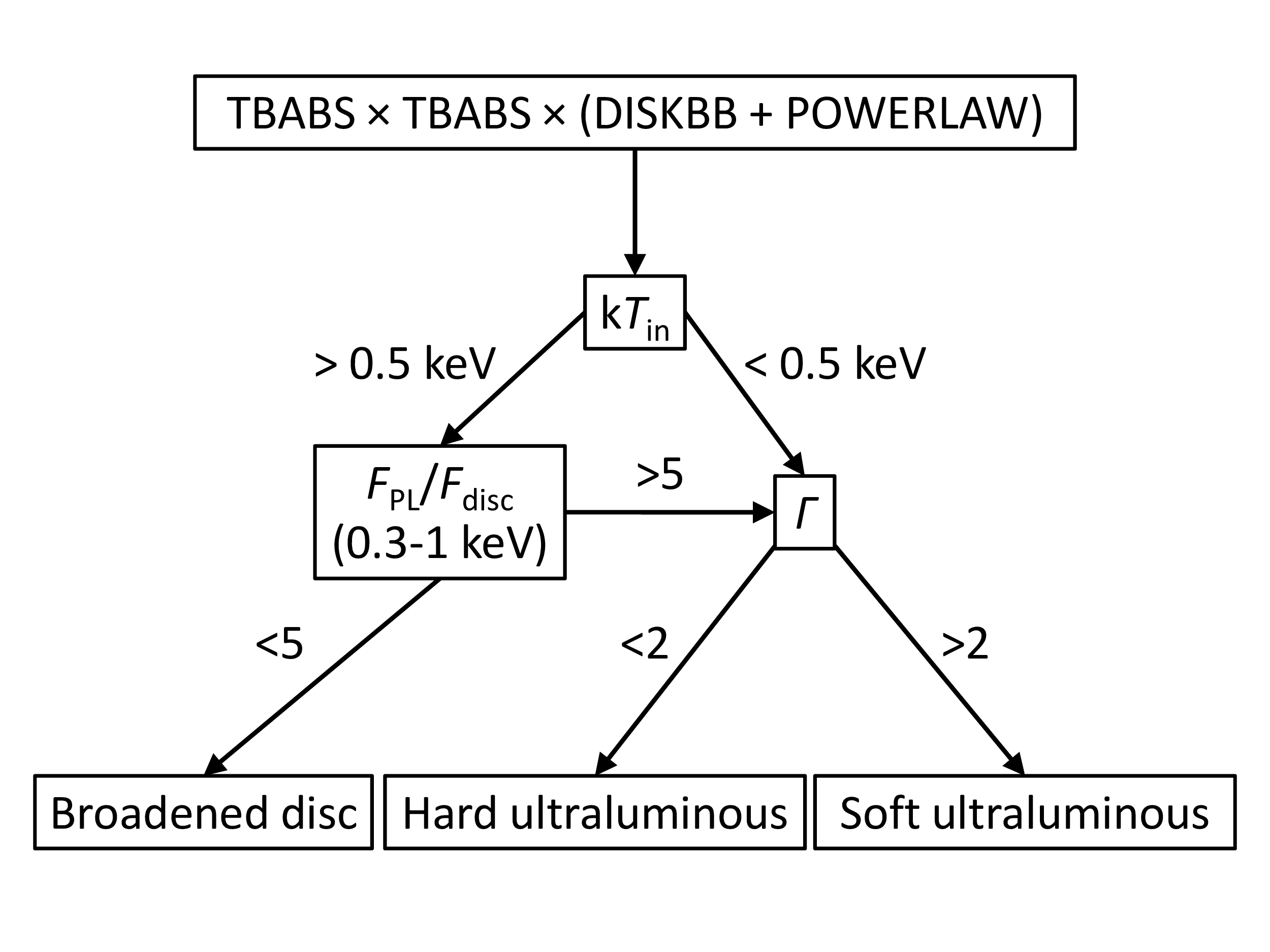}
\caption{Decision tree showing the procedure by which observations were assigned into the three spectral types based on the resulting parameters from the doubly-absorbed MCD plus power-law spectral model.}
\label{decision_tree}
\end{figure}

The first step was to attempt to classify each of the 89 ULX observations into one of the three proposed super-Eddington spectral regimes outlined above and shown in Fig.~\ref{eg_spec}.  We chose to do this using a simple empirical model fit to the data, specifically a doubly-absorbed multi-colour-disc (MCD) plus power-law model ({\sc constant} $\times$ {\sc tbabs} $\times$ {\sc tbabs} $\times$ ({\sc diskbb} + {\sc powerlaw}) in {\sc xspec}).  This is motivated by the work of \cite{gladstone_etal_2009}: in their work the disc-like spectra had relatively warm discs in this model ($kT_{\rm in} > 0.5$ keV), whereas the hard and soft ultraluminous spectra had cool discs and were themselves differentiated by their value of photon index $\Gamma$.  This was $> 2$ for the soft ultraluminous spectra, and $< 2$ for the hard ultraluminous sources.  We therefore base our classification scheme on these simple distinctions; we show the whole scheme in the form of a decision tree in Fig.~\ref{decision_tree}, and discuss it further below.

In order to perform the classification, the energy spectra were grouped to a minimum of 20 counts per energy bin, to allow the spectral fitting to be carried out using $\chi^2$ statistics in {\sc xspec} v12.6.0.   The results of this fitting are shown in Appendix~\ref{appendix} (Table \ref{disc_plus_pow}).  The first absorption component was set equal to the Galactic value in the direction of the galaxy hosting the ULX (as per Table~\ref{sample}), and the latter was left free to model absorption intrinsic to the host galaxy and/or the source itself.  Both components used the interstellar abundance tables of \cite{wilms_etal_2000}.  Additionally, a multiplicative constant was included to account for calibration uncertainties between the detectors, which was fixed to 1 for the PN data (or MOS1 where no PN data was available), and left free to vary for the other detectors.  The constant differed by no more than 10\% between detectors in most cases, with the small number of exceptions being attributable to different extraction regions and/or proximity to the edge of a chip.  We also extracted the 0.3--1 keV absorbed flux contributed by each of the two model components, using the {\sc cflux} convolution model in {\sc xspec}\footnote{To do this it was necessary to rearrange the spectral model such that absorption components were included separately for both the MCD and power-law, which were set to be equal, i.e. {\sc constant} $\times$ ({\sc tbabs} $\times$ {\sc tbabs} $\times$  {\sc diskbb} + {\sc tbabs} $\times$ {\sc tbabs} $\times$ {\sc comptt}), with the {\sc cflux} model component inserted before either the first or third absorption component.}.  The ratio of these fluxes, plus the best fitting disc temperature and the photon index from the MCD plus power-law model were then used as the basis of the empirical classification scheme to diagnose the spectral state, as shown in Fig.~\ref{decision_tree}.  

While the classification scheme is attractive due to its simplicity, it does have limitations.  One of the key diagnostics of the hard and soft ultraluminous spectra is their high energy turnover.  In most cases this feature is rather subtle, so is well-approximated by the power-law component in the MCD plus power-law model (cf. Fig.~\ref{eg_spec}).  However, in ULXs with a stronger break the simple classification scheme can break down, which notably occurs in NGC 55 ULX in the sample of \cite{gladstone_etal_2009}.  In this case the power-law cannot adequately approximate the strong high energy curvature, but a disc spectrum can; the two components switch places as the MCD is forced to the hard end of the spectrum, and the resulting hot disc temperature could then be naively associated with a broadened disc ULX.   However, the strong soft excess in this source is then dominated by a very soft power-law component, so we can break the model degeneracy using the 0.3--1 keV flux ratio of the MCD and power-law by rejecting a broadened disc interpretation where a dominant fraction of the soft emission is in the power-law.  The precise values chosen are motivated by NGC 55 ULX, and we include this modification in the complete classification scheme in Fig.~\ref{decision_tree}.  This solution seems to be sufficient to identify soft ultraluminous spectra with pronounced hard curvature, however it is far from clear how to distinguish hard ultraluminous ULXs with pronounced curvature and little soft excess from broadened discs.  A good example in this work occurs for observation 0145190101 of NGC 5907 ULX, where the best fitting MCD plus power-law model indicates a broadened disc spectrum; however, we reject this interpretation based on a detailed multi-mission study of this source \citep{sutton_etal_2013} and use the parameters from a secondary minimum in $\chi^2$ space to instead classify it as a hard ultraluminous spectrum.  We return to this point later in Section 4.

The example of NGC 5907 ULX leads to another potential source for ambiguity: \cite{sutton_etal_2013} find that the high absorption column in the host galaxy likely suppresses the soft excess emission, such that it is not visible in the spectrum.  Such degeneracy between absorption and the intrinsic spectral shape could again be a limitation of our empirical classification scheme; for example both IC 342 X-1 and X-2 both sit behind a large Galactic absorption column, and in three out of four observations their spectral state is not constrained.  As such, we further identify all highly absorbed ULX observations, and caution that the state identification in these sources is less certain.  All sources with a combined column density consistent with $\ga 0.5 \times 10^{22}~{\rm cm^{-2}}$ in the absorbed MCD plus Comptonisation model (see below) were classed as highly absorbed; this model was used for this purpose in preference to the MCD plus power-law, as a power-law representation of a Comptonised spectrum can become unphysical at low energies and can result in the absorption being over-estimated.  On this basis, high absorption was seen in all observations of IC 342 X-1, IC 342 X-2 and NGC 5907 ULX, plus observation 0200670101 of NGC 6946 X-1.

One further complication arose when fitting the MCD plus power-law to the spectra from NGC 253 ULX2.  This source is embedded in extended emission from the galaxy's disc and an outflow of hot gas from the starburst nucleus (\citealt{pietsch_etal_2001}; \citealt{bauer_etal_2008}), and initially the spectral fits were rather poor.  However, the inclusion of an additional {\sc mekal} component (with abundances frozen to solar values) to model the extended diffuse emission was able to improve this.  As the {\sc mekal} model represents emission from the hot collisionally-ionised gas in the region surrounding the source, it would not be expected to vary between observations.  So, we were able to constrain the spectral parameters of the hot gas by simultaneously fitting all of the observations of NGC 253 ULX2 with an invariable hot gas, plus a variable ULX model.   To constrain the hot gas parameters we again used the more physically motivated ULX model, namely an MCD, plus emission from a Comptonising corona and in this case the diffuse emission from the hot gas ({\sc constant} $\times$ {\sc tbabs} $\times$ ({\sc mekal} + {\sc tbabs} $\times$ ({\sc diskbb} + {\sc comptt})) in {\sc xspec}).  The model parameters of the {\sc mekal} component were set to be equal in each detector for all of the observations, whilst the other model variables\footnote{These were the extra-Galactic absorption column; both the temperature and normalisation of the accretion disc; and the plasma temperature, optical depth and normalisation of the corona (the input soft photon temperature was set equal to the inner disc temperature).} were allowed to vary between observations, but were set to be identical for all of the detectors in a particular observation.  The multiplicative constant was fixed to 1 for the PN detector, and free to vary for the MOS detector in all of the observations.  The resulting best fitting temperature of the hot gas was $kT = 0.67^{+0.10}_{-0.05}$ keV, with a normalisation of $1.5 \pm 0.3 \times 10^{-5} \rm ~cm^{-5}$, which is consistent with the models of the hotter of two plasmas identified in the central region by \cite{bauer_etal_2008}.  Subsequently, we included the {\sc mekal} emission as a fixed additive component in the spectral model for NGC 253 ULX2, both when repeating the empirical state classification and in the following spectral analysis.

The resulting spectral classifications of the ULX sample are shown in Table \ref{disc_plus_pow}.  Out of a total of 89 observations, 43 were uniquely classified as broadened discs, 15 as hard ultraluminous and 21 as soft ultraluminous.  The classification of the remaining 10 observations was ambiguous, as the $1 \sigma$ error bounds of the parameters were consistent with multiple spectral state identifications.

\begin{figure*}
\centering
\includegraphics[width=14.cm, angle=0]{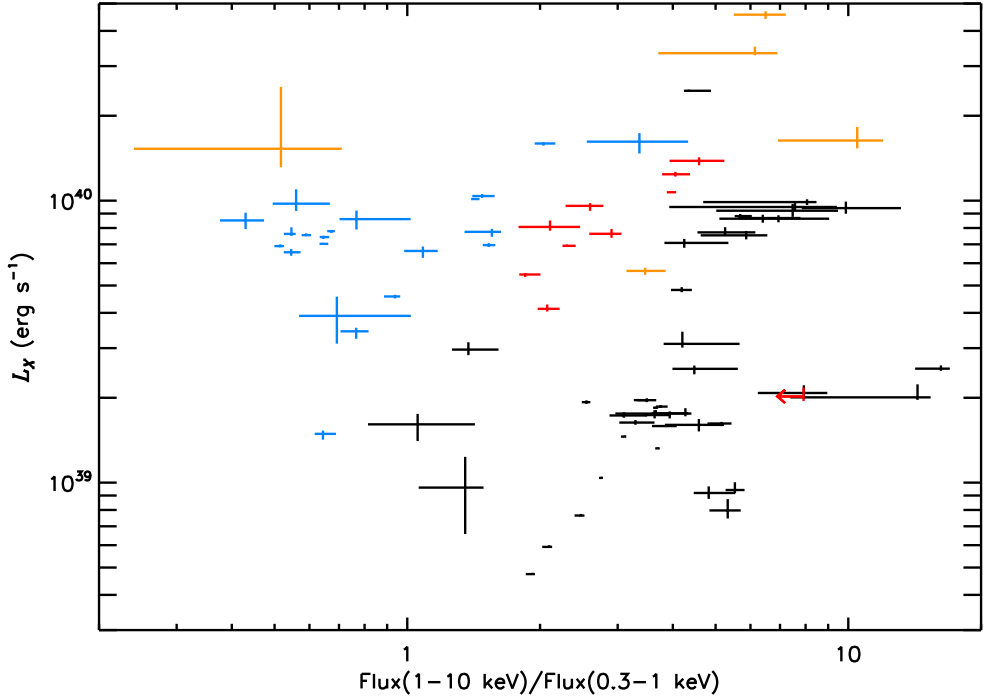}
\caption{Hardness - luminosity diagram for the ULX sample.  Spectral hardness is approximated as the ratio of the unabsorbed 0.3--1 and 1--10 keV fluxes extracted from the absorbed disc plus Comptonising corona model, and is plotted against the 0.3--10 keV unabsorbed luminosity.  The sample is split into different ULX spectral regimes, which are identified by colour: broadened discs in black, hard ultraluminous in red and soft ultraluminous in blue. Only observations with a firm spectral classification are shown.  Also, any highly absorbed observations (with $1 \sigma$ upper limits of $N_{\rm H} \ge 0.5 \times 10^{22}~{\rm cm^{-2}}$) are shown in orange, as the classification scheme is less reliable in highly absorbed sources.   The different spectral types of ULXs tend to occupy different regions of the diagram: few hard ultraluminous or soft ultraluminous sources are seen below $\sim 3 \times 10^{39}~{\rm erg~s^{-1}}$, instead below this luminosity we see mostly modified disc ULXs; the hard and soft ultraluminous sources tend to be brighter, and are generally split by spectral hardness, as would be expected given the method by which we differentiate between them.}
\label{HID}
\end{figure*}

\subsection{Characterising the ULX observations using X-ray luminosity and spectral hardness}\label{mcd_plus_comptt}

As noted above, the use of an MCD plus power-law model can create problems when extracting fluxes that are corrected for absorption given the unphysical behaviour of a power-law at low energies.  Therefore, in order to extract band-limited fluxes from the data, the X-ray spectra of the sample ULXs were also fitted using a second two component spectral model - a more physical absorbed MCD plus a Comptonising corona ({\sc constant} $\times$ {\sc tbabs} $\times$ {\sc tbabs} $\times$ ({\sc diskbb} + {\sc comptt}) in {\sc xspec}).  The additional advantage of this model is that cool, optically thick Comptonisation can better fit the hard spectral curvature seen in many ULXs (cf. \citealt{gladstone_etal_2009}).  As above, a multiplicative constant and two absorption components were included in the spectral model.  The coronal input soft photon temperature was set equal to the inner disc temperature; whilst this is not entirely physical, as we are not necessarily observing the intrinsic inner disc temperature, it does provide a reasonable first approximation (cf. \citealt{pintore_and_zampieri_2012} and the discussion therein).  This model typically resulted in acceptable fits to the majority of the ULX data.  However, the limitations of the model were apparent in a small number of high quality data sets, and it was rejected at $3 \sigma$ significance in 9 of the 89 observations, which notably included the 5 highest data quality observations of NGC 5408 X-1.  A physical interpretation of the best fitting spectral parameters is beyond the scope of the work presented here, and we just used the model to extract X-ray fluxes; we will return to this data, and in particular the features causing the poor fits in future work (Middleton et al. in prep.).  Two observations of IC 342 X-2 had unconstrained disc components and were well fitted by the Comptonisation model alone, so a single-component absorbed Comptonisation model was used for the subsequent analysis of these data.

To characterise the X-ray spectrum seen in each of the observations, we extracted intrinsic (i.e. deabsorbed) fluxes in the full, soft and hard energy bands (0.3--10, 0.3--1 and 1--10 keV), using the {\sc cflux} model component in {\sc xspec} and the MCD plus Comptonising corona model.  Spectral hardness was approximated as the ratio of the hard and soft band intrinsic fluxes, and 0.3--10 keV luminosities were also calculated using the full-band flux and the distances to the host galaxies from Table \ref{sample}; these are shown in the form of a hardness-luminosity diagram in Fig.~\ref{HID}, with the observations split into the three proposed spectral types using our empirical classification method.  Few hard ultraluminous and soft ultraluminous ULXs are seen below $\sim 3 \times 10^{39}~{\rm erg~s^{-1}}$, and the ULXs less luminous than this tend to show broadened discs.  However, a few broadened discs are seen at luminosities greater than $\sim 3 \times 10^{39}~{\rm erg~s^{-1}}$.  As expected, given that the two states are differentiated by the spectral index of the power-law tail, the hard ultraluminous sources tend to be spectrally harder than the soft ultraluminous sources, although they both seem to occupy a similar range of X-ray luminosities, with 12 out of 15 of the hard ultraluminous observations and 20 out of 21 of the soft ultraluminous observations being seen at 0.3--10 keV luminosities between $\sim 0.3$--$2 \times 10^{40}~{\rm erg~s^{-1}}$.

\subsection{State changes in individual ULXs: ambiguous classifications or real spectral progression?}

\begin{figure}
\centering
\includegraphics[width=8.cm, angle=0]{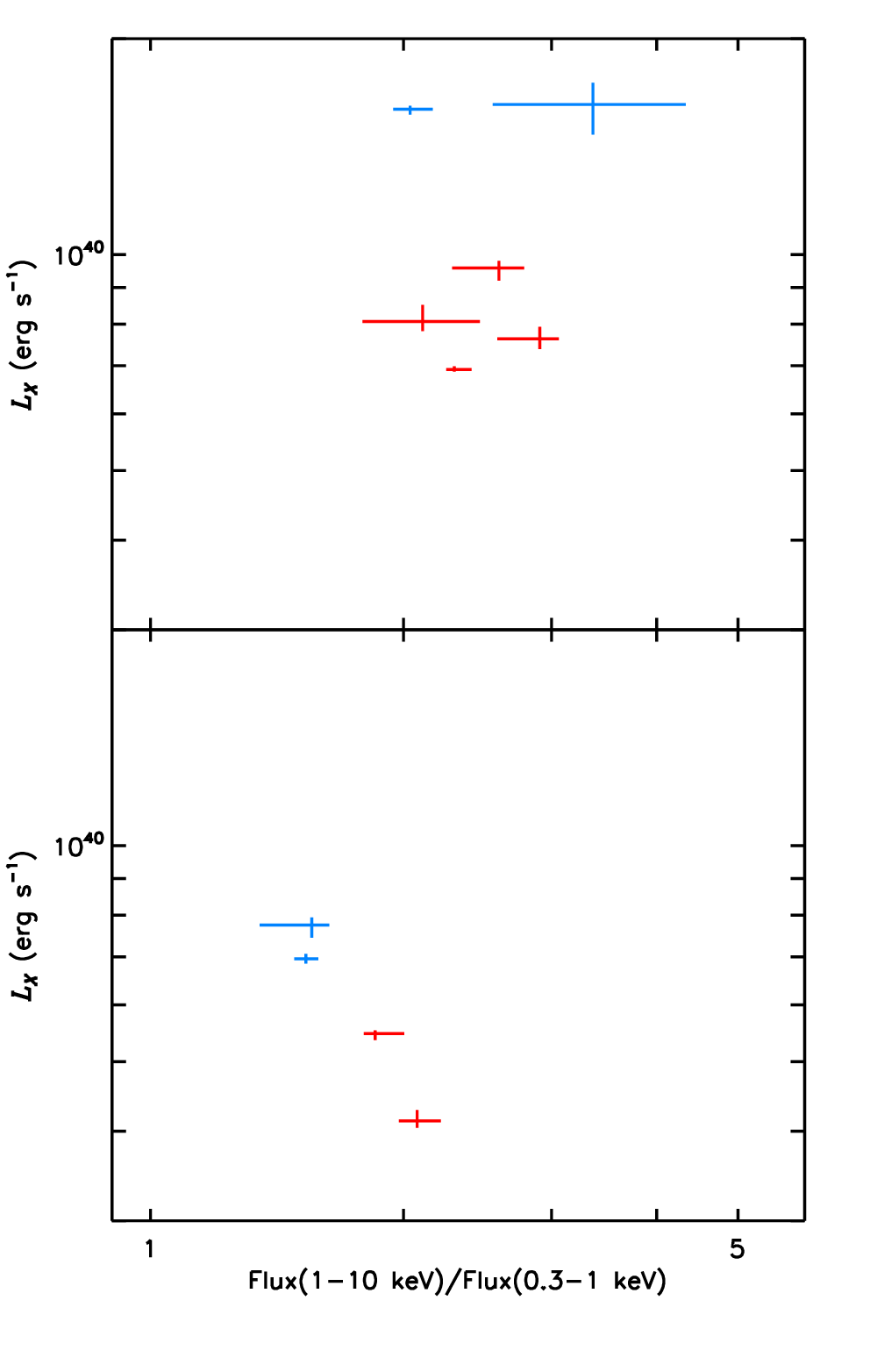}
\caption{Hardness - luminosity diagrams, showing a sub-sample of the data from Figure \ref{HID} for two individual ULXs that are seen as both hard and soft ultraluminous sources; the sources are: ({\it top}) NGC 1313 X-1 and ({\it bottom}) NGC 5204 X-1.  Again, colours represent the different ultraluminous state spectral regimes: red - hard ultraluminous and blue - soft ultraluminous.}
\label{HID2}
\end{figure}

Of the 20 ULXs in the sample, 15 retained the same spectral classification throughout all of the observations included in this study when observations with an ambiguous spectral state are excluded; although 4 of these ULXs only had a single observation with a firm state identification, and the spectral state of IC 342 X-2 was not well constrained in any observation.  However, NGC 253 ULX2, NGC 1313 X-1, Ho IX X-1, NGC 5204 X-1 and NGC 6946 ULX3 were all classified as being in two different spectral states.

NGC 253 ULX2 was observed with a disc-like spectrum in all but one observation, where it appeared with a hard ultraluminous spectrum at a similar deabsorbed 0.3--10 keV luminosity (estimated using an absorbed MCD plus Comptonisation model; see section~\ref{mcd_plus_comptt}).  In this particular case, it seems likely that it may be a case of mistaken identity, possibly due to the relatively low data quality ($\sim$ 1000 counts).  When fitting the MCD plus power-law model to observation 0304851001 of NGC 253 ULX2 the best fitting hard ultraluminous state solution had a fit statistic of $\chi^2/{\rm dof} = 181.4/168$, but there was a second local minima in $\chi^2$ space, with a fit statistic of $\chi^2/{\rm dof}=184.4/168$, which was instead consistent with this observation having a broadened disc spectrum.

Ho IX X-1 was observed with a hard ultraluminous spectrum in three observations, and a broadened disc in one.  Unusually, the broadened disc observation occurred at the highest X-ray luminosity.  In this case, the data quality is not particularly poor, and we suggest that this observation may be misclassified due to the disc-like appearance of the strong spectral curvature in a highly optically-thick Comptonised spectrum (cf. the `very thick' spectral interpretation of NGC 1313 X-1 and X-2 by \citealt{pintore_and_zampieri_2012}).  Again, this highlights the issue in disentangling broadened discs and strongly Comptonised hard ultraluminous regime spectra.

Both hard and soft ultraluminous spectra were identified in observations of NGC 1313 X-1, NGC 5204 X-1 and NGC 6946 X-1.  In the case of NGC 6946 X-1, the location of the two spectral types on the hardness - luminosity diagram appeared to be rather degenerate.  In this case, we suggest that this was unlikely to be due to a real change in the spectral regime, as the spectral indices in the MCD plus power-law fits to NGC 6946 X-1 were all very close to the value of 2 used to separate the states.  However, the hard ultraluminous and soft ultraluminous states in NGC 1313 X-1 and NGC 5204 X-1 do appear to clearly occupy distinct regions of hardness-luminosity space (Fig.~\ref{HID2}); and, the soft ultraluminous spectra were only seen to occur at greater X-ray luminosities than the hard ultraluminous spectra.  Such spectral behaviour as observed in these two ULXs would seem to be consistent with the X-ray luminosity - spectral regime progression as originally suggested by \cite{gladstone_etal_2009}.

\subsection{Short-term X-ray timing analysis}

In addition to the spectral analysis detailed above, we also carried out some basic timing analysis using the diagnostic of fractional variability.  Fractional variability and the associated errors were calculated following the prescription of \cite{vaughan_etal_2003}; it should be noted that this does not account for the effects of intrinsic scattering in the red noise, although \citeauthor{vaughan_etal_2003} state that differences in a fractional variability spectrum that are significantly larger than the estimated uncertainty would indicate achromatic variability.  The main motivation behind this analysis was to provide an additional diagnostic of the spectral state, by e.g. checking for multiple variability components, and to test for consistency with the ultraluminous spectral progression suggested by \cite{gladstone_etal_2009}.  In addition to the full-band (0.3--10 keV) fractional variability, we also calculated values using the energy resolved light curves in a soft (0.3--1 keV) and hard (1--10 keV) energy band.  Ideally we wanted to approximate the relative variability contributions from each of the components in the two component spectrum, but it is not possible to do this precisely, as the components overlap in energy and do not occur over the same energy range in all observations; however, whilst the detection of significantly different hard and soft fractional variabilities would not completely rule out single component spectral models (see e.g., \citealt{gierlinski_and_zdziarski_2005}), it could provide a flag for a likely two component spectrum with different levels of variability in each component.

\begin{figure*}
\centering
\includegraphics[width=14.cm, angle=0]{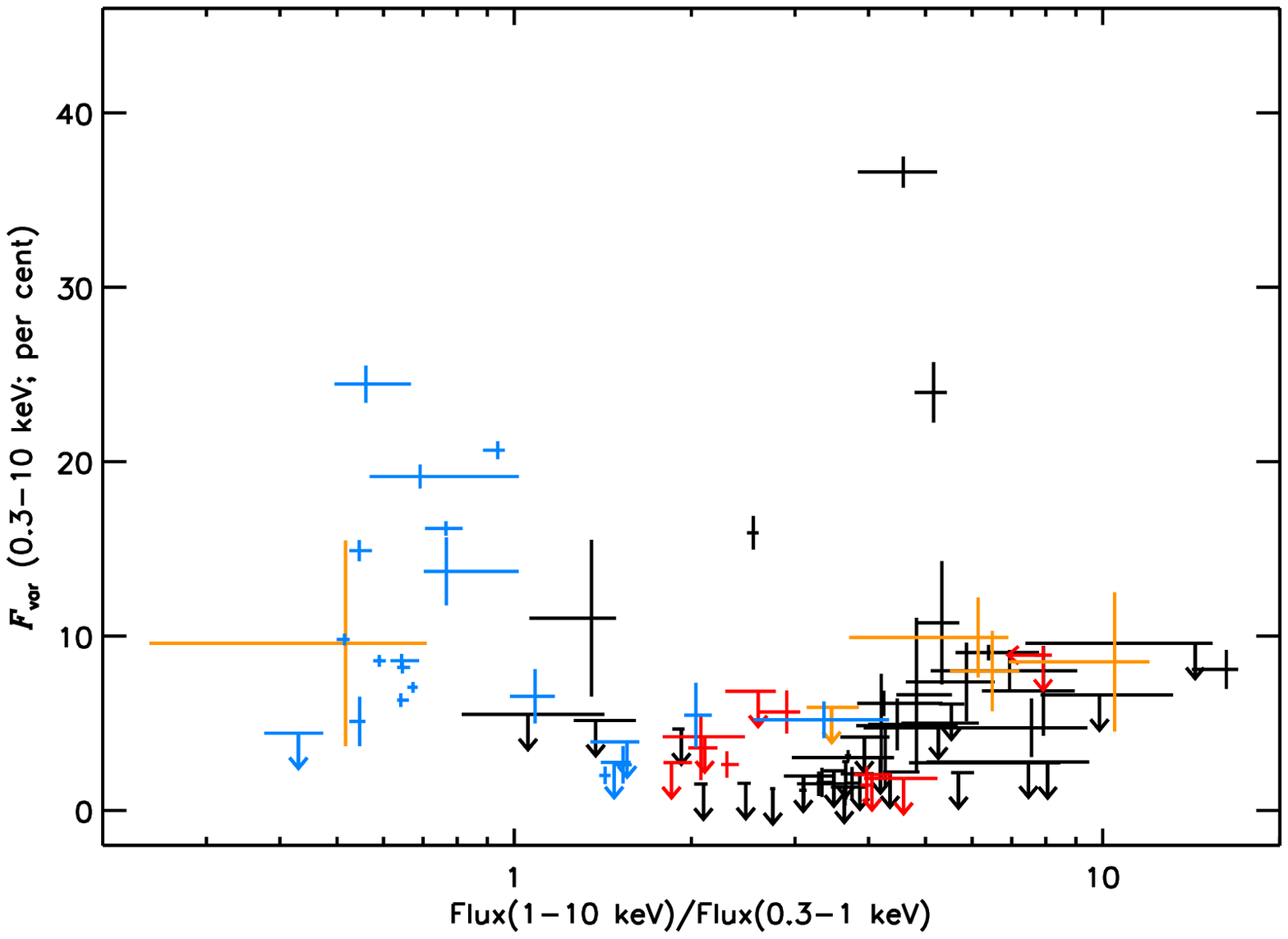}
\includegraphics[width=14.cm, angle=0]{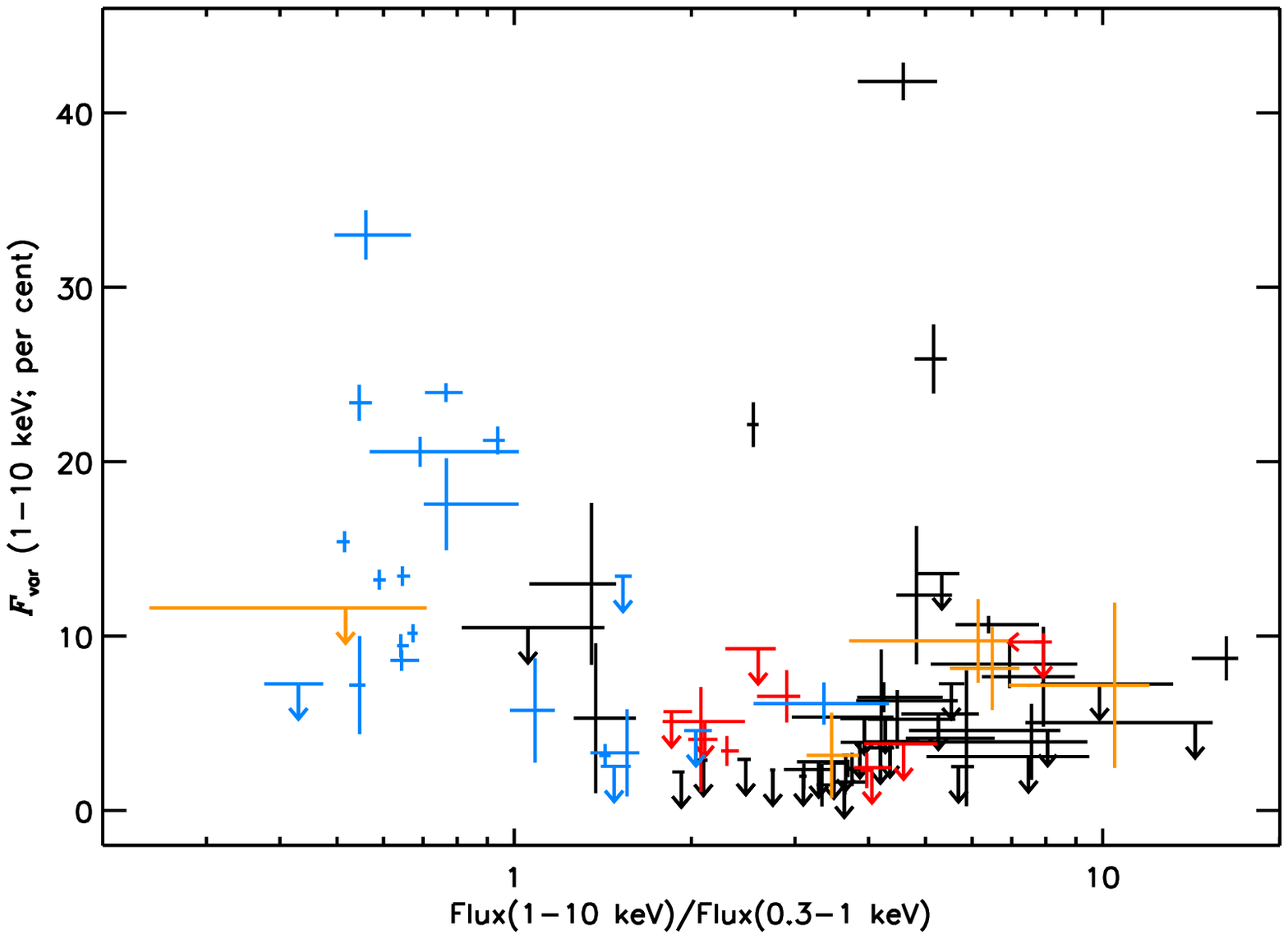}
\caption{Full-band ({\it top}; 0.3--10 keV) and hard-band ({\it bottom}; 1--10 keV) fractional variability of the ULX sample plotted against spectral hardness.  Fractional variability was calculated from light curves binned to 200s.  Errors and limits shown are at the $1 \sigma$ level. The colours correspond to the three ultraluminous spectral regimes, as per previous figures: broadened discs in black, hard ultraluminous in red and soft ultraluminous in blue.  Only observations with a firm spectral classification are shown.  Also, any highly absorbed observations (with $1 \sigma$ upper limits of $N_{\rm H} \ge 0.5 \times 10^{22}~{\rm cm^{-2}}$) are shown in orange, as the classification scheme is less reliable in highly absorbed sources.}
\label{F_var_fig}
\end{figure*}

To calculate fractional variability, we extracted light curves in each of the three energy bands; these were extracted with 200s temporal binning, allowing us to probe variability on time scales from the corresponding Nyquist frequency up to the full available good time in each observation.  The resulting values of fractional variability are shown in Table~\ref{f_var}, and similarly to the hardness-luminosity diagram (Fig.~\ref{HID}) above, we show the full-band fractional variabilities of ULX observations of each spectral type, plotted against spectral hardness (Fig.~\ref{F_var_fig}; {\it top}).  

A range of variability behaviours were detected, and whilst many were consistent with upper limits of only a few per cent fractional variability (cf. \citealt{heil_etal_2009}), this was not the case for all observations.  Variability was detected in all three spectral classes of ULXs, although it was rather weak in the hard ultraluminous observations, where it was limited to at most $\sim 10$ per cent.  However, some soft ultraluminous observations were much more variable, having full-band fractional variabilities of up to $\sim 30$ per cent; there were also a few highly variable detections (up to $\sim 40$ per cent) of disc-like ULXs: observations 0125960101 and 0152020101 of NGC 253 XMM2, plus observation 0404980101 of NGC 4736 ULX1 were all inconsistent (at greater than $3 \sigma$ significance) with having less than 10 per cent 0.3--10 keV fractional variability.  

\begin{figure*}
\centering
\includegraphics[width=14.cm, angle=0]{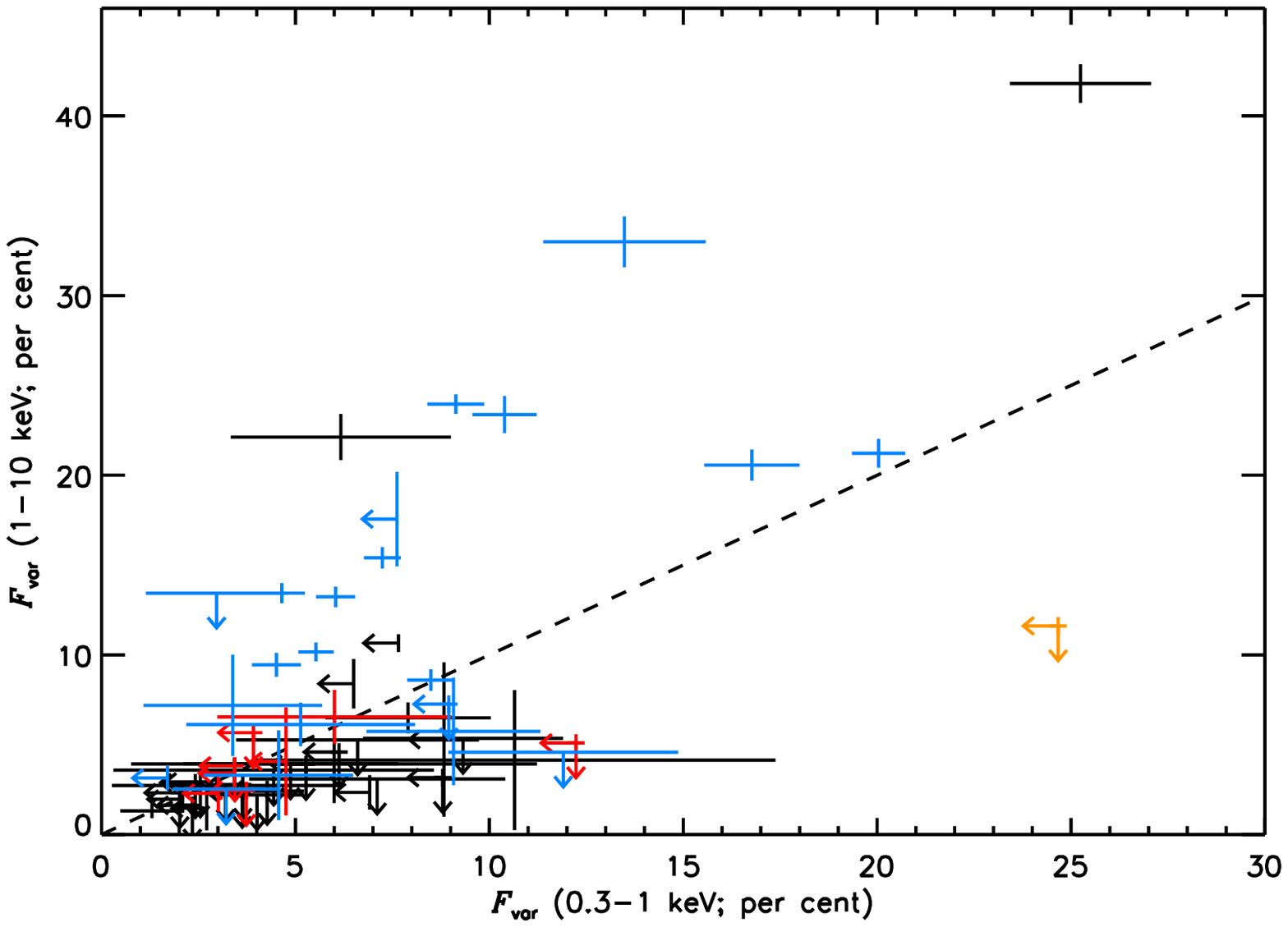}
\caption{Hard-band (1--10 keV) fractional variability plotted against soft-band (0.3--1 keV) fractional variability for the ULX sample.  The dashed line corresponds to equal values of fractional variability in each of the energy bands.  Fractional variabilities were calculated from light curves with 200s temporal binning in both energy bands.  The errors and limits shown are the $1 \sigma$ uncertainty limits, and the colours correspond to the spectral regimes, as per the previous figures.}
\label{F_var_F_var}
\end{figure*}

The high levels of variability in some of the soft ultraluminous observations were even more pronounced in the hard-band (Fig.~\ref{F_var_fig}; {\it bottom}).  Indeed, when we compare the fractional variabilities in the hard and soft energy bands (Fig.~\ref{F_var_F_var}), it was evident that most of the variability detected in these soft ULXs was actually seen in the emission from the hard component.  Intriguingly, the high levels of variability in the hard emission were not always present in the soft ultraluminous observations, rather they seem to be transient.  For example, of the 3 observations of Ho II X-1, all of which were classified as having soft ultraluminous spectra, only 1 showed strong hard variability ($F_{\rm var}$(1--10 keV)$ = 21.2 \pm 0.8$ per cent; compared to $<3$ and $3.1 \pm 0.7$ per cent).  There were a few ULXs with soft ultraluminous observations in which we did not detect strong variability, these were NGC 1313 X-1, NGC 4559 ULX2 and NGC 5204 X-1 (plus IC 342 X-2; although the identification of a soft ultraluminous spectrum in this source is highly questionable).  However, each of these sources was only observed with a soft ultraluminous spectrum in at most 2 epochs, and so had only limited opportunities for the possibly transient variability to be detected.

\section{Discussion}\label{discussion}

We have presented results from an X-ray spectral and timing analysis of a ULX sample, based on the highest quality available archival {\it XMM-Newton} data (plus a proprietary observation of NGC 5907 ULX, obtained for a different study).  Now we attempt to interpret these results, with particular reference to the framework of suggested spectral progression in the super-Eddington ultraluminous state.  We begin by discussing the X-ray properties seen in the sample, before attempting to interpret them in terms of models of super-Eddington accretion.

\subsection{Interpreting the X-ray spectral and timing properties of the ULX sample}

Below $\sim 3 \times 10^{39}~{\rm erg~s^{-1}}$ the ULX population is dominated by observations with broadened disc-like spectra.  Critically, this dominant population is consistent with the suggestion from the \cite{gladstone_etal_2009} that these sources represent accretion rates at around the Eddington limit, with the luminosity limit strongly suggestive of accretion onto sMBHs of $\la 20 \Msun$.  However, disc-like spectra are also observed at higher X-ray luminosities in three objects (NGC 1313 X-2, M81 X-6 and NGC 4190 ULX1), up to $\sim 10^{40}~{\rm erg~s^{-1}}$.  If these sources are indeed powered by $\sim$ Eddington rate accretion, as the population of less-luminous disc-like ULXs implies, then we are either overestimating the distance to these brighter sources, they have a higher degree of beaming than the fainter disc-like sources, or -- most interestingly -- we may be observing the effects of $\sim$ Eddington rate accretion onto the larger MsBHs.  However, it is pertinent here to again note the difficulty in distinguishing between hard ultraluminous spectra with pronounced curvature and broadened disc spectra.  Given the position of these objects at the harder end of the ultraluminous regime objects in the hardness-intensity diagram, and that other ULXs observed at these luminosities appear in the super-Eddington hard ultraluminous or soft ultraluminous regimes, then the mis-identification of strongly Comptonised spectra with little or no soft excess as a broadened disc should be seriously considered.  So, if the apparently disc-like ULXs above $\sim 3 \times 10^{39}~{\rm erg~s^{-1}}$ intrinsically have hard ultraluminous spectra then MsBHs are not required, and at least this sample of ULXs can then be produced solely by $\sim$ Eddington and super-Eddington accretion on to sMBHs.  Clearly, future work to break this degeneracy in the classification scheme is of particular importance.

There are a number of issues in interpreting the broadened-disc ULXs as standard accretion discs.  Not only are they broadened, such that it can be demonstrated that they are not well fit by standard models of disc emission (c.f. \citealt{gladstone_etal_2009}; \citeauthor{middleton_etal_2011b} 2011b; \citealt{middleton_etal_2012}), a few were also highly variable.  This was the case in two observations of NGC 253 XMM2 and one of NGC 4736 ULX1.  There were sufficient statistics in one observation of each source to extract a value of fractional variability in both the soft and hard energy bands; and, in each of these cases the light curve of the hard-band was significantly more variable than the soft-band, as is the case in the soft ultraluminous sources.  If these are indeed disc spectra, then the possible origin of the variability is unclear; and it would seemingly rule out slim disc models.  Rather, the variability properties would suggest that these spectra require a two component solution, in which case the broadened disc ULXs may actually be an emerging hard/soft ultraluminous spectrum.  Indeed, it has been shown that M31 ULX1 (\citealt{kaur_etal_2012}; \citealt{middleton_etal_2012}) and M33 X-8 (\citeauthor{middleton_etal_2011b} 2011b), whilst both disc-like in appearance, are instead better explained as emerging two component hard/soft ultraluminous-like spectra.

The hard and soft ultraluminous regime detections are almost exclusively brighter than $\sim 3 \times 10^{39}~{\rm erg~s^{-1}}$.  In the absence of a highly contrived distribution of black hole masses with these spectral types only occurring around more massive black holes, the lack of fainter sources in either state implies that they must be occurring at luminosities well in excess of the Eddington limit.  Although both the hard and soft ultraluminous regimes are seen to occur over a similar range of X-ray luminosities, the similarities end there.  The hard ultraluminous spectra tend to be spectrally harder than the soft ultraluminous spectra, although this is not particularly surprising given that we differentiate between the two states based on the hardness of a power-law tail.  Less trivial to explain is the difference in variability properties between the two states: variability in the hard ultraluminous state is rather weak and does not appear to exceed $\sim 10$ per cent in any individual object (and is much lower in most), whilst much stronger variability is seen in a number of the soft ultraluminous observations.  The differing variability properties support the distinctions made by our spectral classification system, but more importantly the nature of the variability can give us vital clues to the processes involved.  Whilst \cite{gladstone_etal_2009} argue that the progression to a soft ultraluminous state is characterised by the onset of a dominant spectral contribution from a soft wind, the emergent variability was observed to peak instead in the hard band.  Such energy dependent variability seems to confirm that the spectral solution must also be at least two component in nature; but it is not obvious how this could be produced using reflection models, or the accretion rate driven spectral progression originally suggested by \cite{gladstone_etal_2009}.  In the latter case, the optically-thick corona covering the inner disc would have to conspire to become variable in some (but not all) epochs, at around the same accretion rate that a strong wind begins to dominate the energy spectrum.  This does not appear plausible.  Instead, a mechanism by which ULXs with a soft ultraluminous spectrum display intermittent, strong variability of the hard component is required.

\subsection{A unified model of ULX accretion}

From the available evidence presented above, it seems clear that the X-ray spectral and timing characteristics of ULXs in the hard ultraluminous and soft ultraluminous regimes cannot be uniquely determined by accretion rate alone.  Rather, we suggest that the key implication of this study is that it favours a model in which both regimes are produced in a single type of system, with the observed characteristics of the source being dependent on both accretion rate and -- critically -- the inclination of the ULX system (cf. \citealt{poutanen_etal_2007}).  The required angular dependence can naturally be introduced into the ULX system by the presence of a massive outflowing, radiatively driven wind.  Such a wind is expected to occur at super-Eddington accretion rates (\citealt{poutanen_etal_2007}; \citealt{dotan_and_shaviv_2011}; \citealt{ohsuga_and_mineshige_2011}), indeed the super-critical accretion flow simulations of \cite{kawashima_etal_2012} produce such a wind, and find that it takes the form of a funnel.  Then, when we observe the ULX system face-on, we see down the funnel and observe emission from the innermost regions (that have a hard, cool corona/hot disc spectrum); however, at inclinations closer to the funnel opening angle, a much softer spectrum from the wind's photosphere is observed.  So, at inclination angles between these, the ULX would appear with a shifting balance in the relative flux contributed from either component.  In this model, the observed association between a soft spectrum and potential variability can also be achieved, if the variability can only occur at particular observation angles where the balance of component fluxes has shifted towards the wind.  This is the case if the edge of the wind is clumpy in nature (cf. \citeauthor{middleton_etal_2011a} 2011a; \citealt{takeuchi_etal_2013}); then, at angles where the wind's edge intersects the line-of-sight to the central source, the clumpiness can result in variable obscuration of the hard emission region, thus extrinsically imprinting the observed, predominantly hard variability.

We argue that a unified ultraluminous state model can produce the spectral timing and properties seen in a large sample of ULXs, but for this model to be believable it is also essential that it can predict the spectral progression within individual sources.  It has been predicted that an increase in the accretion rate onto a ULX would result in a narrowing of the wind's opening angle \citep{king_2008}, thus shifting the emission from the various components relative to a fixed line-of-sight.  Then, a ULX could potentially transit from the hard ultraluminous state to the soft ultraluminous state with increasing accretion rate, as the wind shifts towards the observer's line-of-sight.  Indeed, we observe such a spectral change in both NGC 1313 X-1 and NGC 5204 X-1.  Further, we would expect to see the onset of the characteristic hard variability as the accretion rate increases still further, and the clumpy edge of the wind enters the line-of-sight; future observations of a ULX progressing through the hard ultraluminous, invariable soft ultraluminous and variable soft ultraluminous regimes could provide a strong test of this model.  

The wind itself is expected to be highly ionised; we might expect to see signatures of this in the form of absorption features in ULX spectra.  \cite{walton_etal_2012} attempted to detect these using the stacked {\it XMM-Newton} spectra of NGC 1313 X-1 and Ho IX X-1, but found no evidence of absorption lines.  However, both of these ULXs have been predominantly observed with hard ultraluminous spectra, which this work argues are likely to have clear lines-of-sight to the central emission regions, so no absorption from material in the wind is expected.  Instead, the model favoured here would predict the detection of the imprint of wind absorption in soft ultraluminous ULXs; this may occur in the higher energy spectrum from the central region, or possibly across the full {\it XMM-Newton} band-pass if the emission from the wind is self-absorbed by the outflowing turbulent material as it becomes more diffuse and optically thin further from the source.  Thus, these ULXs may display some observational similarities to high mass X-ray binaries, that are seen through columns of $> 10^{23}~{\rm cm^{-2}}$ of ionised stellar wind (e.g. \citealt{torrejon_etal_2010}).  We will revisit this in future work (Middleton et al. in prep.).

It is also interesting to consider the multi-wavelength properties of ULXs in each regime.  Radio nebulae have been identified around two ULXs in the soft ultraluminous state - Ho II X-1 and NGC 5408 X-1 (\citealt{kaaret_etal_2003}; \citealt{soria_etal_2006}; \citealt{lang_etal_2007}; \citealt{miller_etal_2005}).  However, radio nebulae are not a unique feature of the soft ultraluminous state, one is also observed around IC 342 X-1 \citep{cseh_etal_2012}, which is in the hard ultraluminous state; if these nebulae are inflated by disc winds, then it may further support the argument that the distinction between hard and soft ultraluminous spectra is largely due to inclination.

\section{Conclusions}

We have presented results from an {\it XMM-Newton} spectral and timing study of a sample of 20 ULXs with some of the highest quality X-ray data available.  Using a new classification scheme, we separated the ULXs into three spectral types based on the results of \cite{gladstone_etal_2009}.  A number of trends were identified in the spectral and timing properties of the sample, which we interpret in terms of a physical model.

\noindent $\bullet$ Below $\sim 3 \times 10^{39}~{\rm erg~s^{-1}}$ the ULX population is dominated by broadened disc spectra.  This is consistent with a population of sMBHs ($M \la 20 M_{\odot}$) accreting at close to, and just above, the Eddington limit.  We therefore suggest that the broadened disc ULXs bridge the luminosity gap between the standard sub-Eddington black hole binaries, and the super-Eddington ULXs with more extreme spectra.  Hence, the brighter ULXs that display mainly hard ultraluminous and soft ultraluminous spectra must be at super-Eddington luminosities.  

\noindent $\bullet$ A few broadened disc ULXs were seen at higher X-ray luminosities.  Rather excitingly, these could be powered by accretion onto larger black hole primaries (MsBHs).   However, more mundane explanations such as higher beaming factors cannot be discounted.    Alternatively, they may be misclassified objects; a hard ultraluminous spectrum with little soft excess could readily be confused with a broadened disc, in which case typical sMBHs may be sufficient to produce the entire population of ULXs up to $\sim 2 \times 10^{40} \rm ~erg~s^{-1}$.

\noindent $\bullet$ Interestingly, a few broadened disc ULXs display strong variability.  Instead of being intrinsically disc-like, this suggests that broadened discs may in fact be emerging two-component hard/soft ultraluminous spectra.

\noindent $\bullet$ The brighter hard ultraluminous and soft ultraluminous ULXs differ by definition in spectral hardness, but crucially also differ in their variability properties.  Low levels of variability are seen in the hard ultraluminous state (fractional variability $\ll 10$ per cent), whilst some soft ultraluminous observations are highly variable (10--30 percent fractional variability), with the variability being strongest in the higher energy part of the spectrum ($> 1$ keV).

\noindent $\bullet$ The observed X-ray properties can be achieved in a model with a funnel shaped wind, as is expected to emerge in super-Eddington sources.  Viewed down the opening angle of the wind we observe hard emission from the central source, and so a geometrically-beamed hard ultraluminous spectrum.  At higher inclinations to our line-of-sight a wind-dominated soft ultraluminous spectrum is seen.  

\noindent $\bullet$ In this model, the observed hard variability in spectrally soft sources originates from the changing obscuration of the hard central emission when observed through the clumpy edge of the out-flowing wind.

\noindent $\bullet$ If, as predicted, the opening angle of the wind narrows with increasing accretion rate, the observed spectrum can change from hard ultraluminous to soft ultraluminous at higher X-ray luminosities.  This is the behaviour we see in NGC 1313 X-1 and NGC 5204 X-1.

This model is predicated on both theory and observation, and goes a long way towards unifying and explaining the X-ray characteristics of ULXs known to us from the best quality data available.  But, as with any model, it requires further testing to confirm our current understanding, and develop deeper knowledge.  We suggest that the key to this will be obtaining more high quality ULX observations that constrain the spectral and timing properties of ULXs to high precision, particularly in objects that are seen to transit between different spectral regimes.  However, until such observations are taken, the data presented in this paper is suggesting to us that the properties of ULXs may be explained solely by a population of sMBHs, with their different observational characteristics dependent upon two variables: their accretion rate, and their inclination angle to our line-of-sight.

\section*{Acknowledgements}

We thank the anonymous referee for their helpful comments.
ADS gratefully acknowledges funding from the Science and Technology Facilities Council in the form of a PhD studentship (ST/F007299/1), and TPR and MJM in the form of a standard grant (ST/G000158/1)).  MJM also acknowledges a Marie Curie FP7 Postdoctoral scholarship.  This work is based on observations obtained with {\it XMM-Newton\/}, an ESA science mission with instruments and contributions directly funded by ESA Member States and NASA; it has also made use of data from the High Energy Astrophysics Science Archive Research Center (HEASARC), which is a service of the Astrophysics Science Division at NASA/GSFC and the High Energy Astrophysics Division of the Smithsonian Astrophysical Observatory. 

\bibliography{refs}

\begin{thebibliography}{}

\bibitem[\protect\citeauthoryear{{Abramowicz}, {Czerny}, {Lasota} \&
  {Szuszkiewicz}}{{Abramowicz} et~al.}{1988}]{abramowicz_etal_1988}
{Abramowicz} M.~A.,  {Czerny} B.,  {Lasota} J.~P.,    {Szuszkiewicz} E.,  1988,
  \apj, 332, 646

\bibitem[\protect\citeauthoryear{{Bauer}, {Pietsch}, {Trinchieri},
  {Breitschwerdt}, {Ehle}, {Freyberg} \& {Read}}{{Bauer}
  et~al.}{2008}]{bauer_etal_2008}
{Bauer} M.,  {Pietsch} W.,  {Trinchieri} G.,  {Breitschwerdt} D.,  {Ehle} M.,
  {Freyberg} M.~J.,    {Read} A.~M.,  2008, \aap, 489, 1029

\bibitem[\protect\citeauthoryear{{Belczynski}, {Bulik}, {Fryer}, {Ruiter},
  {Valsecchi}, {Vink} \& {Hurley}}{{Belczynski}
  et~al.}{2010}]{belczynski_etal_2010}
{Belczynski} K.,  {Bulik} T.,  {Fryer} C.~L.,  {Ruiter} A.,  {Valsecchi} F.,
  {Vink} J.~S.,    {Hurley} J.~R.,  2010, \apj, 714, 1217

\bibitem[\protect\citeauthoryear{{Caballero-Garc{\'{\i}}a} \&
  {Fabian}}{{Caballero-Garc{\'{\i}}a} \&
  {Fabian}}{2010}]{caballero_garcia_and_fabian_2010}
{Caballero-Garc{\'{\i}}a} M.~D.,  {Fabian} A.~C.,  2010, \mnras, 402, 2559

\bibitem[\protect\citeauthoryear{{Colbert} \& {Mushotzky}}{{Colbert} \&
  {Mushotzky}}{1999}]{colbert_and_mushotzky_1999}
{Colbert} E.~J.~M.,  {Mushotzky} R.~F.,  1999, \apj, 519, 89

\bibitem[\protect\citeauthoryear{{Cseh}, {Corbel}, {Kaaret}, {Lang},
  {Gris{\'e}}, {Paragi}, {Tzioumis}, {Tudose} \& {Feng}}{{Cseh}
  et~al.}{2012}]{cseh_etal_2012}
{Cseh} D.,  {Corbel} S.,  {Kaaret} P.,  {Lang} C.,  {Gris{\'e}} F.,  {Paragi}
  Z.,  {Tzioumis} A.,  {Tudose} V.,    {Feng} H.,  2012, \apj, 749, 17

\bibitem[\protect\citeauthoryear{{Dalcanton} et~al.,}{{Dalcanton}
  et~al.}{2009}]{dalcanton_etal_2009}
{Dalcanton} J.~J.,  et~al., 2009, \apjs, 183, 67

\bibitem[\protect\citeauthoryear{{de Vaucouleurs}, {de Vaucouleurs}, {Corwin}
  Jr., {Buta}, {Paturel} \& {Fouque}}{{de Vaucouleurs}
  et~al.}{1991}]{deVaucouleurs_etal_1991}
{de Vaucouleurs} G.,  {de Vaucouleurs} A.,  {Corwin} Jr. H.~G.,  {Buta} R.~J.,
  {Paturel} G.,    {Fouque} P.,  1991, {Third Reference Catalogue of Bright
  Galaxies, Vols 1--3}.
Springer-Verlag, Berlin, Heidelberg, New York, p.~7

\bibitem[\protect\citeauthoryear{{Dickey} \& {Lockman}}{{Dickey} \&
  {Lockman}}{1990}]{dickey_and_lockman_1990}
{Dickey} J.~M.,  {Lockman} F.~J.,  1990, \araa, 28, 215

\bibitem[\protect\citeauthoryear{{Done}, {Wardzi{\'n}ski} \&
  {Gierli{\'n}ski}}{{Done} et~al.}{2004}]{done_etal_2004}
{Done} C.,  {Wardzi{\'n}ski} G.,    {Gierli{\'n}ski} M.,  2004, \mnras, 349,
  393

\bibitem[\protect\citeauthoryear{{Dotan} \& {Shaviv}}{{Dotan} \&
  {Shaviv}}{2011}]{dotan_and_shaviv_2011}
{Dotan} C.,  {Shaviv} N.~J.,  2011, \mnras, 413, 1623

\bibitem[\protect\citeauthoryear{{Farrell}, {Webb}, {Barret}, {Godet} \&
  {Rodrigues}}{{Farrell} et~al.}{2009}]{farrell_etal_2009}
{Farrell} S.~A.,  {Webb} N.~A.,  {Barret} D.,  {Godet} O.,    {Rodrigues}
  J.~M.,  2009, \nat, 460, 73

\bibitem[\protect\citeauthoryear{{Feng} \& {Soria}}{{Feng} \&
  {Soria}}{2011}]{feng_and_soria_2011}
{Feng} H.,  {Soria} R.,  2011, New Astron. Rev., 55, 166

\bibitem[\protect\citeauthoryear{{Gierli{\'n}ski} \&
  {Zdziarski}}{{Gierli{\'n}ski} \&
  {Zdziarski}}{2005}]{gierlinski_and_zdziarski_2005}
{Gierli{\'n}ski} M.,  {Zdziarski} A.~A.,  2005, \mnras, 363, 1349

\bibitem[\protect\citeauthoryear{{Gladstone}, {Roberts} \& {Done}}{{Gladstone}
  et~al.}{2009}]{gladstone_etal_2009}
{Gladstone} J.~C.,  {Roberts} T.~P.,    {Done} C.,  2009, \mnras, 397, 1836

\bibitem[\protect\citeauthoryear{{Gon{\c c}alves} \& {Soria}}{{Gon{\c c}alves}
  \& {Soria}}{2006}]{goncalves_and_soria_2006}
{Gon{\c c}alves} A.~C.,  {Soria} R.,  2006, \mnras, 371, 673

\bibitem[\protect\citeauthoryear{{Heil}, {Vaughan} \& {Roberts}}{{Heil}
  et~al.}{2009}]{heil_etal_2009}
{Heil} L.~M.,  {Vaughan} S.,    {Roberts} T.~P.,  2009, \mnras, 397, 1061

\bibitem[\protect\citeauthoryear{{Henze}, {Pietsch}, {Haberl} \&
  {Greiner}}{{Henze} et~al.}{2009}]{henze_etal_2009}
{Henze} M.,  {Pietsch} W.,  {Haberl} F.,    {Greiner} J.,  2009, The
  Astronomer's Telegram, 2356, 1

\bibitem[\protect\citeauthoryear{{Herrmann}, {Ciardullo}, {Feldmeier} \&
  {Vinciguerra}}{{Herrmann} et~al.}{2008}]{herrmann_etal_2008}
{Herrmann} K.~A.,  {Ciardullo} R.,  {Feldmeier} J.~J.,    {Vinciguerra} M.,
  2008, \apj, 683, 630

\bibitem[\protect\citeauthoryear{{Jacobs}, {Rizzi}, {Tully}, {Shaya}, {Makarov}
  \& {Makarova}}{{Jacobs} et~al.}{2009}]{EDD2}
{Jacobs} B.~A.,  {Rizzi} L.,  {Tully} R.~B.,  {Shaya} E.~J.,  {Makarov} D.~I.,
    {Makarova} L.,  2009, \aj, 138, 332

\bibitem[\protect\citeauthoryear{{Kaaret}, {Corbel}, {Prestwich} \&
  {Zezas}}{{Kaaret} et~al.}{2003}]{kaaret_etal_2003}
{Kaaret} P.,  {Corbel} S.,  {Prestwich} A.~H.,    {Zezas} A.,  2003, Science,
  299, 365

\bibitem[\protect\citeauthoryear{{Kajava} \& {Poutanen}}{{Kajava} \&
  {Poutanen}}{2009}]{kajava_and_poutanen_2009}
{Kajava} J.~J.~E.,  {Poutanen} J.,  2009, \mnras, 398, 1450

\bibitem[\protect\citeauthoryear{{Kajava}, {Poutanen}, {Farrell}, {Gris{\'e}}
  \& {Kaaret}}{{Kajava} et~al.}{2012}]{kajava_etal_2012}
{Kajava} J.~J.~E.,  {Poutanen} J.,  {Farrell} S.~A.,  {Gris{\'e}} F.,
  {Kaaret} P.,  2012, \mnras, 422, 990

\bibitem[\protect\citeauthoryear{{Karachentsev} et~al.,}{{Karachentsev}
  et~al.}{2002}]{karachentsev_etal_2002}
{Karachentsev} I.~D.,  et~al., 2002, \aap, 385, 21

\bibitem[\protect\citeauthoryear{{Karachentsev} et~al.,}{{Karachentsev}
  et~al.}{2003}]{karachentsev_etal_2003}
{Karachentsev} I.~D.,  et~al., 2003, \aap, 398, 467

\bibitem[\protect\citeauthoryear{{Karachentsev}, {Sharina} \&
  {Huchtmeier}}{{Karachentsev} et~al.}{2000}]{karachentsev_etal_2000}
{Karachentsev} I.~D.,  {Sharina} M.~E.,    {Huchtmeier} W.~K.,  2000, \aap,
  362, 544

\bibitem[\protect\citeauthoryear{{Kaur}, {Henze}, {Haberl}, {Pietsch},
  {Greiner}, {Rau}, {Hartmann}, {Sala} \& {Hernanz}}{{Kaur}
  et~al.}{2012}]{kaur_etal_2012}
{Kaur} A.,  {Henze} M.,  {Haberl} F.,  {Pietsch} W.,  {Greiner} J.,  {Rau} A.,
  {Hartmann} D.~H.,  {Sala} G.,    {Hernanz} M.,  2012, \aap, 538, A49

\bibitem[\protect\citeauthoryear{{Kawashima}, {Ohsuga}, {Mineshige}, {Yoshida},
  {Heinzeller} \& {Matsumoto}}{{Kawashima} et~al.}{2012}]{kawashima_etal_2012}
{Kawashima} T.,  {Ohsuga} K.,  {Mineshige} S.,  {Yoshida} T.,  {Heinzeller} D.,
     {Matsumoto} R.,  2012, \apj, 752, 18

\bibitem[\protect\citeauthoryear{{King}}{{King}}{2004}]{king_2004}
{King} A.~R.,  2004, \mnras, 347, L18

\bibitem[\protect\citeauthoryear{{King}}{{King}}{2008}]{king_2008}
{King} A.~R.,  2008, \mnras, 385, L113

\bibitem[\protect\citeauthoryear{{Lang}, {Kaaret}, {Corbel} \& {Mercer}}{{Lang}
  et~al.}{2007}]{lang_etal_2007}
{Lang} C.~C.,  {Kaaret} P.,  {Corbel} S.,    {Mercer} A.,  2007, \apj, 666, 79

\bibitem[\protect\citeauthoryear{{Liu} \& {Bregman}}{{Liu} \&
  {Bregman}}{2005}]{liu_and_bregman_2005}
{Liu} J.-F.,  {Bregman} J.~N.,  2005, \apjs, 157, 59

\bibitem[\protect\citeauthoryear{{Mapelli}, {Ripamonti}, {Zampieri}, {Colpi} \&
  {Bressan}}{{Mapelli} et~al.}{2010}]{mapelli_etal_2010}
{Mapelli} M.,  {Ripamonti} E.,  {Zampieri} L.,  {Colpi} M.,    {Bressan} A.,
  2010, \mnras, 408, 234

\bibitem[\protect\citeauthoryear{{Middleton}, {Done}, {Ward}, {Gierli{\'n}ski}
  \& {Schurch}}{{Middleton} et~al.}{2009}]{middleton_etal_2009}
{Middleton} M.,  {Done} C.,  {Ward} M.,  {Gierli{\'n}ski} M.,    {Schurch} N.,
  2009, \mnras, 394, 250

\bibitem[\protect\citeauthoryear{{Middleton}, {Roberts}, {Done} \&
  {Jackson}}{{Middleton} et~al.}{011a}]{middleton_etal_2011a}
{Middleton} M.~J.,  {Roberts} T.~P.,  {Done} C.,    {Jackson} F.~E.,  2011a,
  \mnras, 411, 644

\bibitem[\protect\citeauthoryear{{Middleton}, {Sutton} \&
  {Roberts}}{{Middleton} et~al.}{011b}]{middleton_etal_2011b}
{Middleton} M.~J.,  {Sutton} A.~D.,    {Roberts} T.~P.,  2011b, \mnras, 417,
  464

\bibitem[\protect\citeauthoryear{{Middleton}, {Sutton}, {Roberts}, {Jackson} \&
  {Done}}{{Middleton} et~al.}{2012}]{middleton_etal_2012}
{Middleton} M.~J.,  {Sutton} A.~D.,  {Roberts} T.~P.,  {Jackson} F.~E.,
  {Done} C.,  2012, \mnras, 420, 2969

\bibitem[\protect\citeauthoryear{{Miller}, {Mushotzky} \& {Neff}}{{Miller}
  et~al.}{2005}]{miller_etal_2005}
{Miller} N.~A.,  {Mushotzky} R.~F.,    {Neff} S.~G.,  2005, \apjl, 623, L109

\bibitem[\protect\citeauthoryear{{Mineo}, {Gilfanov} \& {Sunyaev}}{{Mineo}
  et~al.}{2012}]{mineo_etal_2012}
{Mineo} S.,  {Gilfanov} M.,    {Sunyaev} R.,  2012, \mnras, 419, 2095

\bibitem[\protect\citeauthoryear{{Miyawaki}, {Makishima}, {Yamada}, {Gandhi},
  {Mizuno}, {Kubota}, {Tsuru} \& {Matsumoto}}{{Miyawaki}
  et~al.}{2009}]{miyawaki_etal_2009}
{Miyawaki} R.,  {Makishima} K.,  {Yamada} S.,  {Gandhi} P.,  {Mizuno} T.,
  {Kubota} A.,  {Tsuru} T.~G.,    {Matsumoto} H.,  2009, \pasj, 61, 263

\bibitem[\protect\citeauthoryear{{Ohsuga} \& {Mineshige}}{{Ohsuga} \&
  {Mineshige}}{2011}]{ohsuga_and_mineshige_2011}
{Ohsuga} K.,  {Mineshige} S.,  2011, \apj, 736, 2

\bibitem[\protect\citeauthoryear{{Pietsch}, {Roberts}, {Sako}, {Freyberg},
  {Read}, {Borozdin}, {Branduardi-Raymont}, {Cappi}, {Ehle}, {Ferrando},
  {Kahn}, {Ponman}, {Ptak}, {Shirey} \& {Ward}}{{Pietsch}
  et~al.}{2001}]{pietsch_etal_2001}
{Pietsch} W.,  {Roberts} T.~P.,  {Sako} M.,  {Freyberg} M.~J.,  {Read} A.~M.,
  {Borozdin} K.~N.,  {Branduardi-Raymont} G.,  {Cappi} M.,  {Ehle} M.,
  {Ferrando} P.,  {Kahn} S.~M.,  {Ponman} T.~J.,  {Ptak} A.,  {Shirey} R.~E.,
   {Ward} M.,  2001, \aap, 365, L174

\bibitem[\protect\citeauthoryear{{Pintore} \& {Zampieri}}{{Pintore} \&
  {Zampieri}}{2012}]{pintore_and_zampieri_2012}
{Pintore} F.,  {Zampieri} L.,  2012, \mnras, 420, 1107

\bibitem[\protect\citeauthoryear{{Poutanen}, {Lipunova}, {Fabrika}, {Butkevich}
  \& {Abolmasov}}{{Poutanen} et~al.}{2007}]{poutanen_etal_2007}
{Poutanen} J.,  {Lipunova} G.,  {Fabrika} S.,  {Butkevich} A.~G.,
  {Abolmasov} P.,  2007, \mnras, 377, 1187

\bibitem[\protect\citeauthoryear{{Rao}, {Feng} \& {Kaaret}}{{Rao}
  et~al.}{2010}]{rao_etal_2010}
{Rao} F.,  {Feng} H.,    {Kaaret} P.,  2010, \apj, 722, 620

\bibitem[\protect\citeauthoryear{{Roberts}}{{Roberts}}{2007}]{roberts_2007}
{Roberts} T.~P.,  2007, \apss, 311, 203

\bibitem[\protect\citeauthoryear{{Roberts}, {Warwick}, {Ward}, {Goad} \&
  {Jenkins}}{{Roberts} et~al.}{2005}]{roberts_etal_2005}
{Roberts} T.~P.,  {Warwick} R.~S.,  {Ward} M.~J.,  {Goad} M.~R.,    {Jenkins}
  L.~P.,  2005, \mnras, 357, 1363

\bibitem[\protect\citeauthoryear{{Soria}}{{Soria}}{2007}]{soria_2007}
{Soria} R.,  2007, \apss, 311, 213

\bibitem[\protect\citeauthoryear{{Soria}, {Fender}, {Hannikainen}, {Read} \&
  {Stevens}}{{Soria} et~al.}{2006}]{soria_etal_2006}
{Soria} R.,  {Fender} R.~P.,  {Hannikainen} D.~C.,  {Read} A.~M.,    {Stevens}
  I.~R.,  2006, \mnras, 368, 1527

\bibitem[\protect\citeauthoryear{{Stobbart}, {Roberts} \& {Wilms}}{{Stobbart}
  et~al.}{2006}]{stobbart_etal_2006}
{Stobbart} A.-M.,  {Roberts} T.~P.,    {Wilms} J.,  2006, \mnras, 368, 397

\bibitem[\protect\citeauthoryear{{Strohmayer} \& {Mushotzky}}{{Strohmayer} \&
  {Mushotzky}}{2003}]{strohmayer_and_mushotzky_2003}
{Strohmayer} T.~E.,  {Mushotzky} R.~F.,  2003, \apjl, 586, L61

\bibitem[\protect\citeauthoryear{{Strohmayer} \& {Mushotzky}}{{Strohmayer} \&
  {Mushotzky}}{2009}]{strohmayer_and_mushotzky_2009}
{Strohmayer} T.~E.,  {Mushotzky} R.~F.,  2009, \apj, 703, 1386

\bibitem[\protect\citeauthoryear{{Strohmayer}, {Mushotzky}, {Winter}, {Soria},
  {Uttley} \& {Cropper}}{{Strohmayer} et~al.}{2007}]{strohmayer_etal_2007}
{Strohmayer} T.~E.,  {Mushotzky} R.~F.,  {Winter} L.,  {Soria} R.,  {Uttley}
  P.,    {Cropper} M.,  2007, \apj, 660, 580

\bibitem[\protect\citeauthoryear{{Sutton}, {Roberts}, {Gladstone}, {Farrell},
  {Reilly}, {Goad} \& {Gehrels}}{{Sutton} et~al.}{2013}]{sutton_etal_2013}
{Sutton} A.~D.,  {Roberts} T.~P.,  {Gladstone} J.~C.,  {Farrell} S.~A.,
  {Reilly} E.,  {Goad} M.~R.,    {Gehrels} N.,  2013, \mnras, {in press
  (arXiv:1306.4825)}

\bibitem[\protect\citeauthoryear{{Sutton}, {Roberts}, {Walton}, {Gladstone} \&
  {Scott}}{{Sutton} et~al.}{2012}]{sutton_etal_2012}
{Sutton} A.~D.,  {Roberts} T.~P.,  {Walton} D.~J.,  {Gladstone} J.~C.,
  {Scott} A.~E.,  2012, \mnras, 423, 1154

\bibitem[\protect\citeauthoryear{{Swartz}, {Ghosh}, {Tennant} \& {Wu}}{{Swartz}
  et~al.}{2004}]{swartz_etal_2004}
{Swartz} D.~A.,  {Ghosh} K.~K.,  {Tennant} A.~F.,    {Wu} K.,  2004, \apjs,
  154, 519

\bibitem[\protect\citeauthoryear{{Takeuchi}, {Ohsuga} \&
  {Mineshige}}{{Takeuchi} et~al.}{2013}]{takeuchi_etal_2013}
{Takeuchi} S.,  {Ohsuga} K.,    {Mineshige} S.,  2013, PASJ, accepted
  (arXiv:1305.1023)

\bibitem[\protect\citeauthoryear{{Tikhonov} \& {Karachentsev}}{{Tikhonov} \&
  {Karachentsev}}{1998}]{tikhonov_and_karachentsev_1998}
{Tikhonov} N.~A.,  {Karachentsev} I.~D.,  1998, \aaps, 128, 325

\bibitem[\protect\citeauthoryear{{Torrej{\'o}n}, {Schulz}, {Nowak} \&
  {Kallman}}{{Torrej{\'o}n} et~al.}{2010}]{torrejon_etal_2010}
{Torrej{\'o}n} J.~M.,  {Schulz} N.~S.,  {Nowak} M.~A.,    {Kallman} T.~R.,
  2010, \apj, 715, 947

\bibitem[\protect\citeauthoryear{{Tully}, {Rizzi}, {Shaya}, {Courtois},
  {Makarov} \& {Jacobs}}{{Tully} et~al.}{2009}]{EDD}
{Tully} R.~B.,  {Rizzi} L.,  {Shaya} E.~J.,  {Courtois} H.~M.,  {Makarov}
  D.~I.,    {Jacobs} B.~A.,  2009, \aj, 138, 323

\bibitem[\protect\citeauthoryear{{Ueda}, {Yamaoka} \& {Remillard}}{{Ueda}
  et~al.}{2009}]{ueda_etal_2009}
{Ueda} Y.,  {Yamaoka} K.,    {Remillard} R.,  2009, \apj, 695, 888

\bibitem[\protect\citeauthoryear{{Vaughan}, {Edelson}, {Warwick} \&
  {Uttley}}{{Vaughan} et~al.}{2003}]{vaughan_etal_2003}
{Vaughan} S.,  {Edelson} R.,  {Warwick} R.~S.,    {Uttley} P.,  2003, \mnras,
  345, 1271

\bibitem[\protect\citeauthoryear{{Vierdayanti}, {Mineshige}, {Ebisawa} \&
  {Kawaguchi}}{{Vierdayanti} et~al.}{2006}]{vierdayanti_etal_2006}
{Vierdayanti} K.,  {Mineshige} S.,  {Ebisawa} K.,    {Kawaguchi} T.,  2006,
  \pasj, 58, 915

\bibitem[\protect\citeauthoryear{{Walton}, {Miller}, {Reis} \&
  {Fabian}}{{Walton} et~al.}{2012}]{walton_etal_2012}
{Walton} D.~J.,  {Miller} J.~M.,  {Reis} R.~C.,    {Fabian} A.~C.,  2012,
  \mnras, 426, 473

\bibitem[\protect\citeauthoryear{{Walton}, {Roberts}, {Mateos} \&
  {Heard}}{{Walton} et~al.}{2011}]{walton_etal_2011b}
{Walton} D.~J.,  {Roberts} T.~P.,  {Mateos} S.,    {Heard} V.,  2011, \mnras,
  416, 1844

\bibitem[\protect\citeauthoryear{{Watson} et~al.,}{{Watson}
  et~al.}{2009}]{watson_etal_2009}
{Watson} M.~G.,  et~al., 2009, \aap, 493, 339

\bibitem[\protect\citeauthoryear{{Wilms}, {Allen} \& {McCray}}{{Wilms}
  et~al.}{2000}]{wilms_etal_2000}
{Wilms} J.,  {Allen} A.,    {McCray} R.,  2000, \apj, 542, 914

\bibitem[\protect\citeauthoryear{{Zampieri} \& {Roberts}}{{Zampieri} \&
  {Roberts}}{2009}]{zampieri_and_roberts_2009}
{Zampieri} L.,  {Roberts} T.~P.,  2009, \mnras, 400, 677

\bibitem[\protect\citeauthoryear{{Zdziarski}, {Grove}, {Poutanen}, {Rao} \&
  {Vadawale}}{{Zdziarski} et~al.}{2001}]{zdziarski_etal_2001}
{Zdziarski} A.~A.,  {Grove} J.~E.,  {Poutanen} J.,  {Rao} A.~R.,    {Vadawale}
  S.~V.,  2001, \apjl, 554, L45

\end{thebibliography}
\bibliographystyle{mn2e}

\appendix
\section[]{ULX X-ray spectral and timing results}\label{appendix}

Here we present the parameters resulting from the MCD plus power-law spectral model fits (Table \ref{disc_plus_pow}), and fractional variability observbed in each observation (Table \ref{f_var}).


\begin{table*}
\caption{Spectral parameters - {\sc tbabs} $\times$ {\sc tbabs} $\times$ ({\sc diskbb} + {\sc power-law})}
\begin{center}
\begin{tabular}{cccccccc}
\hline
Obs. ID$^a$ & $\chi^2/{\rm dof} ^b$ & ${N_{\rm H}}^c$ & ${kT_{\rm in}}^d$ & $\Gamma^e$ & ${F_{\rm PL}/F_{\rm disc}} ^f$ & ${L_{\rm X}}^g$ & Spectral regime$^h$ \\
& & & & & (0.3--1 keV) & (0.3--10 keV) & \\
\hline
\multicolumn{8}{c}{NGC 55 ULX} \\

0028740201 & 928.9/881 & $0.49 \pm 0.01$ & $0.81 \pm 0.02$ & $3.80^{+0.10}_{-0.09}$ & $14 \pm 1$ & $3.4^{+0.1}_{-0.2}$ & SUL\\
0028740101 & 690.2/669 & $0.45^{+0.03}_{-0.02}$ & $0.87 \pm 0.04$ & $3.4 \pm 0.2$ & $9 \pm 1$ & $3.9^{+0.7}_{-0.8}$ & SUL\\
0655050101 & 1084.9/865 & $0.54 \pm 0.01$ & $0.67 \pm 0.02$ & $4.3 \pm 0.08$ & $28 \pm 3$ & $1.49^{+0.04}_{-0.07}$ & SUL \\
\\
\multicolumn{8}{c}{M31 ULX1} \\

0600660201 & 1147.5/1042 & $0.07 \pm 0.01$ & $1.09 \pm 0.01$ & $2.7 \pm 0.1$ & $0.95 \pm 0.08$ & $1.457^{+0.008}_{-0.007}$ & disc \\
0600660301 & 940.7/1002 & $0.080^{+0.017}_{-0.009}$ & $1.025^{+0.014}_{-0.009}$ & $2.79^{+0.17}_{-0.09}$ & $1.26^{+0.08}_{-0.09}$ & $1.040 \pm 0.005$ & disc \\
0600660401 & 995.7/977 & $0.09^{+0.02}_{-0.01}$ & $0.947^{+0.012}_{-0.008}$ & $3.0^{+0.2}_{-0.1}$ & $1.22^{+0.09}_{-0.12}$ & $0.765^{+0.010}_{-0.006}$ & disc \\
0600660501 & 948.8/900 & $0.11^{+0.02}_{-0.01}$ & $0.86 \pm 0.01$ & $3.0^{+0.2}_{-0.1}$ & $1.6 \pm 0.1$ & $0.592^{+0.008}_{-0.005}$ & disc \\
0600660601 & 871.0/857 & $0.10^{+0.02}_{-0.01}$ & $0.80 \pm 0.01$ & $3.0 \pm 0.1$ & $1.6^{+0.2}_{-0.1}$ & $0.474^{+0.005}_{-0.003}$ & disc \\ 
\\
\multicolumn{8}{c}{NGC 253 XMM2} \\

0125960101 & 192.7/180 & $0.08^{+0.31}_{-0.04}$ & $1.11^{+0.34}_{-0.07}$ & $1 \pm 3$ & $<0.8$ & $1.62^{+0.02}_{-0.03}$ & disc \\
0110900101 & 47.9/44 & $<0.06$ & $0.50^{+0.06}_{-0.05}$ & $0.8^{+0.4}_{-0.9}$ & $1.0^{+0.1}_{-0.2}$ & $0.42^{+0.02}_{-0.03}$ & disc/HUL \\
0152020101 & 532.6/541 & $0.15^{+0.03}_{-0.02}$ & $1.23^{+0.09}_{-0.07}$ & $2.2^{+0.3}_{-0.2}$ & $1.6^{+0.3}_{-0.2}$ & $1.60 \pm 0.08$ & disc \\
0304850901 & 102.6/97 & $<0.3$ & $1.4 \pm 0.1$ & $3^{+1}_{-6}$ & $<2$ & $0.80^{+0.08}_{-0.05}$ & disc \\
0304851001 & 115.5/110 & $0.08 \pm 0.05$ & $1.1^{+0.4}_{-0.2}$ & $1.5 \pm 0.4$ & $<2$ & $0.92^{+0.05}_{-0.04}$ & disc \\
0304851201 & 197.6/190 & $0.17^{+0.10}_{-0.08}$ & $1.40^{+0.08}_{-0.07}$ & $2.9^{+0.9}_{-1.3}$ & $0.9^{+0.3}_{-0.5}$ & $0.84^{+0.06}_{-0.02}$ & disc \\
\\
\multicolumn{8}{c}{NGC 253 ULX2$^i$} \\

0125960101 & 597.9/581 & $0.21^{+0.03}_{-0.02}$ & $1.52 \pm 0.03$ & $>5$ & $0.06^{+0.07}_{-0.02}$ & $2.54^{+0.07}_{-0.04}$ & disc \\
0110900101 & 166.8/160 & $0.28^{+0.04}_{-0.03}$ & $1.64^{+0.07}_{-0.05}$ & $>6$ & $0.15^{+0.09}_{-0.06}$ & $3.11^{+0.33}_{-0.09}$ & disc \\
0152020101 & 614.3/624 & $0.33^{+0.08}_{-0.05}$ & $1.69^{+0.05}_{-0.14}$ & $2.2^{+0.8}_{-0.9}$ & $0.4 \pm 0.2$ & $2.53^{+0.07}_{-0.11}$ & disc \\
0304850901 & 179.4/176 & $0.26^{+0.03}_{-0.02}$ & $1.59 \pm 0.05$ & - & $0.11 \pm 0.07$ & $2.1 \pm 0.1$ & disc \\
0304851001 & 181.4/168 & $0.56 \pm 0.05$ & $0.035^{+0.006}_{-0.007}$ & $1.79^{+0.03}_{-0.05}$ & $3.1 \pm 0.6$ & $2.02^{+0.15}_{-0.07}$ & HUL \\
0304851201 & 319.6/287 & $0.25^{+0.12}_{-0.04}$ & $1.52 \pm 0.04$ & $>4$ & $0.2^{+0.3}_{-0.1}$ & $2.01^{+0.22}_{-0.04}$ & disc \\
0304851301 & 42.5/45 & $0.29^{+0.06}_{-0.05}$ & $1.4^{+0.2}_{-0.1}$ & $<10$ & $<1$ & $1.0 \pm 0.3$ & disc \\
\\
\multicolumn{8}{c}{M33 X-8} \\

0102640401 & 360.4/364 & $0.16^{+0.05}_{-0.03}$ & $1.13 \pm 0.08$ & $2.4^{+0.4}_{-0.2}$ & $2.0 \pm 0.3$ & $1.84 \pm 0.01$ & disc \\
0102640101 & 1022.9/1035 & $0.10^{+0.01}_{-0.02}$ & $1.06^{+0.02}_{-0.04}$ & $2.08^{+0.07}_{-0.12}$ & $1.2 \pm 0.1$ & $1.63 \pm 0.03$ & disc \\
0102640701 & 360.3/350 & $0.16 \pm 0.02$ & $1.0 \pm 0.2$ & $2.1^{+0.2}_{-0.1}$ & $5^{+1}_{-2}$ & $1.79 \pm 0.04$ & disc/HUL/SUL \\
0102641001 & 346.1/315 & $0.17^{+0.04}_{-0.03}$ & $1.4 \pm 0.1$ & $2.4^{+0.3}_{-0.2}$ & $4.0^{+0.7}_{-0.6}$ & $1.76^{+0.08}_{-0.04}$ & disc \\
0102642001 & 689.0/698 & $0.11 \pm 0.02$ & $0.89 \pm 0.05$ & $1.9^{+0.1}_{-0.2}$ & $1.5^{+0.4}_{-0.5}$ & $1.96 \pm 0.03$ & disc \\
0102642101 & 1078.7/1015 & $0.08^{+0.02}_{-0.01}$ & $1.01^{+0.04}_{-0.03}$ & $1.91^{+0.11}_{-0.09}$ & $1.0 \pm 0.1$ & $1.86 \pm 0.02$ & disc \\
0102642301 & 1060.9/1049 & $0.14^{+0.02}_{-0.01}$ & $1.10 \pm 0.03$ & $2.33^{+0.12}_{-0.09}$ & $1.6^{+0.1}_{-0.2}$ & $1.96 \pm 0.02$ & disc \\
0141980601 & 842.3/813 & $0.09 \pm 0.02$ & $0.99^{+0.06}_{-0.04}$ & $1.8^{+0.2}_{-0.1}$ & $1.1 \pm 0.3$ & $1.73^{+0.05}_{-0.03}$ & disc \\
0141980801 & 1163.8/1018 & $0.11 \pm 0.01$ & $0.88^{+0.05}_{-0.04}$ & $1.94^{+0.05}_{-0.06}$ & $3.0 \pm 0.5$ & $1.74^{+0.04}_{-0.05}$ & disc \\
0141980101 & 331.7/308 & $0.14^{+0.04}_{-0.03}$ & $0.97^{+0.12}_{-0.08}$ & $2.1^{+0.3}_{-0.2}$ & $2.1^{+0.4}_{-1.0}$ & $1.76^{+0.03}_{-0.07}$ & disc \\

\hline
\end{tabular}
\end{center}
\begin{minipage}{\linewidth}
Here we report the model parameters, and the resulting spectral classification from fitting the ULX observations with a doubly-absorbed MCD plus power-law model.  Errors and limits are shown at the $1 \sigma$ level, and dashes indicate an unconstrained parameter.
$^a${\it XMM-Newton} observation identifiers.
$^b$Statistical goodness of fit, in terms of of $\chi^2$ and the number of degrees of freedom.
$^c$Extragalactic absorption column density ($\times 10^{22}~{\rm cm^{-2}}$).
$^d$Inner disc temperature in keV.
$^e$Spectral index of the power-law component.
$^f$The ratio of the observed 0.3--1 keV component fluxes from the power-law and MCD.
$^g$0.3--10 keV deabsorbed luminosity ($\times 10^{39}~{\rm erg~s^{-1}}$), calculated from the absorbed MCD plus Comptonisation model.
$^h$The resulting spectral classification of the observation, these are either broadened disc (disc), hard ultraluminous (HUL) or soft ultraluminous (SUL).
$^i$An additional {\sc mekal} component was included in the spectral model of NGC 253 ULX2 to model extended emission in the host galaxy; this had parameters fixed to the best fitting values from simultaneously fitting all observations with an appropriate model ({\sc tbabs} $\times$ ({\sc mekal} + {\sc tbabs} $\times$ ({\sc diskbb} + {\sc comptt})), with a constant {\sc mekal} component, but the parameters of the ULX model free to vary between observations.
$^j$The fit reported for observation 0145190101 of NGC 5907 ULX is a local minimum in $\chi^2$ space, with $\chi^2/{\rm dof}=204.5/168$.  The global minimum occurs for a $2.1^{+0.2}_{-0.1}$ keV hot disc at $\chi^2/{\rm dof}=184.5/168$, which would result in the source being classified as a broadened disc during this observation.  However, the combination of a hot, low mass black hole and the extremely high luminosity of the source would require an unusually large Eddington ratio of $\sim 100$, which would seem rather unphysical.  The very high line-of-sight absorption column of NGC 5907 ULX, likely due to its location in an edge on spiral galaxy, may suppress the soft excess, resulting in the ambiguity in spectral classification.
\end{minipage}
\label{disc_plus_pow}
\end{table*}

\begin{table*}
\begin{flushleft} 
{\bf Table \ref{disc_plus_pow}.} (continued)
\end{flushleft}
\begin{center}
\begin{tabular}{cccccccc}
\hline
Obs. ID$^a$ & $\chi^2/{\rm dof} ^b$ &${N_{\rm H}}^c$ & ${kT_{\rm in}}^d$ & $\Gamma^e$ & ${F_{\rm PL}/F_{\rm disc}} ^f$ & ${L_{\rm X}}^g$ & Spectral regime$^h$ \\
 & & & & & (0.3--1 keV) & (0.3--10 keV)& \\
\hline
\multicolumn{8}{c}{M33 X-8 (continued)} \\
0141980301 & 539.2/500 & $0.06 \pm 0.02$ & $0.96^{+0.05}_{-0.03}$ & $1.9^{+0.2}_{-0.1}$ & $0.9^{+0.2}_{-0.3}$ & $1.59^{+0.02}_{-0.01}$ & disc \\
0650510101 & 1964.8/1658 & $0.10^{+0.05}_{-0.08}$ & $0.99^{+0.01}_{-0.02}$ & $1.98^{+0.03}_{-0.05}$ & $1.18 \pm 0.07$ & $1.322 \pm 0.006$ & disc \\
0650510201 & 2422.2/1967 & $0.090 \pm 0.005$ & $1.05 \pm 0.01$ & $2.06 \pm 0.03$ & $1.08 \pm 0.05$ & $1.789^{+0.006}_{-0.005}$ & disc \\
\\
\multicolumn{8}{c}{NGC 1313 X-1} \\

0106860101 & 443.6/473 & $0.21 \pm 0.02$ & $0.26 \pm 0.03$ & $1.69 \pm 0.05$ & $1.8 \pm 0.4$ & $7.6^{+0.3}_{-0.2}$ & HUL \\
0150280301 & 271.3/254 & $0.39^{+0.04}_{-0.03}$ & $2^{+2}_{-1}$ & $2.5^{+0.3}_{-0.2}$ & $>20$ & $15.9^{+0.2}_{-0.3}$ & SUL \\
0150280601 & 323.6/327 & $0.27 \pm 0.03$ & $0.38^{+0.05}_{-0.06}$ & $1.9^{+0.1}_{-0.2}$ & $1.7^{+0.5}_{-0.6}$ & $11.2^{+2.8}_{-0.5}$ & HUL/SUL \\
0205230301 & 574.6/601 & $0.44^{+0.03}_{-0.04}$ & $0.08^{+0.01}_{-0.03}$ & $2.35^{+0.03}_{-0.04}$ & $12^{+7}_{-9}$ & $16 \pm 1$ & SUL \\
0205230501 & 224.9/222 & $0.25 \pm 0.03$ & $0.29^{+0.05}_{-0.04}$ & $1.75^{+0.09}_{-0.10}$ & $1.5 \pm 0.3$ & $9.6^{+0.2}_{-0.4}$ & HUL \\
0205230601 & 219.8/218 & $0.27 \pm 0.04$ & $0.23^{+0.04}_{-0.03}$ & $1.72^{+0.07}_{-0.08}$ & $1.3 \pm 0.3$ & $8.1^{+0.4}_{-0.2}$ & HUL \\
0405090101 & 1657.9/1457 & $0.26 \pm 0.01$ & $0.227^{+0.009}_{-0.008}$ & $1.68 \pm 0.02$ & $1.33^{+0.07}_{-0.05}$ & $6.92^{+0.07}_{-0.05}$ & HUL \\
\\
\multicolumn{8}{c}{NGC 1313 X-2} \\

0106860101 & 167.6/185 & $0.39^{+0.05}_{-0.06}$ & $2.0^{+0.9}_{-0.2}$ & $3.2^{+0.3}_{-0.5}$  & $3 \pm 1$ & $3.0^{+0.2}_{-0.1}$ & disc \\
0150280301 & 412.2/423 & $0.22 \pm 0.03$ & $1.6 \pm 0.3$ & $1.7 \pm 0.2$ & $2.8^{+0.7}_{-1.0}$ & $8.6^{+0.2}_{-0.3}$ & disc \\
0150280601 & 183.1/193 & $0.27 \pm 0.04$ & $1.0^{+0.4}_{-0.3}$ & $2.3^{+0.1}_{-0.2}$ & $11^{+6}_{-10}$ & $3.3^{+0.2}_{-0.3}$ & disc/HUL \\
0205230301 & 546.4/580 & $0.29^{+0.04}_{-0.03}$ & $1.9^{+0.1}_{-0.2}$ & $2.1^{+0.4}_{-0.2}$ & $2.5^{+0.4}_{-0.3}$ & $9.9 \pm 0.2$ & disc \\
0205230501 & 282.2/280 & $0.36 \pm 0.04$ & $2.1^{+0.5}_{-0.3}$ & $3.0 \pm 0.3$ & $4 \pm 1$ & $3.0 \pm 0.1$ & disc/SUL \\
0205230601 & 344.2/356 & $0.30^{+0.12}_{-0.06}$ & $2.1 \pm 0.1$ & $2.4^{+0.9}_{-0.5}$ & $2.1 \pm 0.3$ & $9.5 \pm 0.3$ & disc \\
0301860101 & 735.1/732 & $0.29^{+0.06}_{-0.03}$ & $1.76^{+0.08}_{-0.06}$ & $2.3^{+0.5}_{-0.2}$ & $1.6 \pm 0.2$ & $9.2 \pm 0.5$ & disc \\
0405090101 & 1402.0/1484 & $0.27^{+0.02}_{-0.01}$ & $1.71^{+0.11}_{-0.08}$ & $1.93^{+0.13}_{-0.07}$ & $2.9 \pm 0.2$ & $8.7^{+0.2}_{-0.3}$ & disc \\

\\
\multicolumn{8}{c}{IC 342 X-1} \\

0093640901 & 145.3/163 & $0.53^{+0.15}_{-0.09}$ & $0.4^{+0.2}_{-0.1}$ & $1.55^{+0.09}_{-0.11}$ & $4 \pm 3$ & $4.4 \pm 0.5$ & disc/HUL \\
0206890201 & 482.0/486 & $0.66^{+0.07}_{-0.06}$ & $0.33^{+0.06}_{-0.05}$ & $1.66 \pm 0.03$ & $1.8 \pm 0.4$ & $5.6^{+0.1}_{-0.2}$ & HUL \\
\\
\multicolumn{8}{c}{IC 342 X-2} \\

0093640901 & 48.9/49 & $2.3^{+0.4}_{-0.2}$ & $2.1 \pm 0.2$ & $>-3$ & $8^{+4}_{-5}$ & $3.7 \pm 0.2$ & disc/HUL/SUL \\
0206890201 & 344.7/389 & $1.8^{+0.4}_{-0.1}$ & $3.2^{+0.1}_{-0.3}$ & $8^{+2}_{-1}$ & $3^{+2}_{-1}$ & $3.73 \pm 0.08$ & disc/SUL \\
\\
\multicolumn{8}{c}{NGC 2403 X-1} \\

0164560901 & 344.6/333 & $0.11^{+0.04}_{-0.05}$ & $1.15^{+0.10}_{-0.09}$ & $2.0^{+0.5}_{-0.3}$ & $<1$ & $1.6^{+0.1}_{-0.2}$ & disc \\
\\
\multicolumn{8}{c}{Ho II X-1} \\

0112520601 & 637.9/684 & $0.11 \pm 0.01$ & $0.36 \pm 0.02$ & $2.38 \pm 0.07$ & $3.3^{+1.1}_{-0.7}$ & $10.4^{+0.2}_{-0.1}$ & SUL \\
0200470101 & 1244.2/1118 & $0.106^{+0.006}_{-0.005}$ & $0.34 \pm 0.01$ & $2.36 \pm 0.03$ & $3.0^{+0.5}_{-0.2}$ & $10.13 \pm 0.06$ & SUL \\
0561580401 & 846.6/753 & $0.098 \pm 0.006$ & $0.21 \pm 0.01$ & $2.46 \pm 0.04$ & $2.4 \pm 0.3$ & $4.57^{+0.06}_{-0.07}$ & SUL \\
\\
\multicolumn{8}{c}{M81 X-6} \\

0111800101 & 1018.8/955 & $0.31 \pm 0.03$ & $1.42 \pm 0.02$ & $2.8^{+0.3}_{-0.2}$ & $1.7^{+0.2}_{-0.1}$ & $7.1^{+0.2}_{-0.3}$ & disc \\
0112521001 & 222.3/200 & $0.33^{+0.14}_{-0.07}$ & $1.9^{+0.2}_{-0.1}$ & $2.6^{+1.0}_{-0.6}$ & $2.2^{+0.7}_{-0.5}$ & $7.7^{+0.3}_{-0.2}$ & disc \\
0112521101 & 191.2/194 & $0.20^{+0.10}_{-0.05}$ & $1.4^{+0.4}_{-0.2}$ & $1.6^{+1.4}_{-0.5}$ & $1.1^{+0.5}_{-0.8}$ & $7.5^{+0.3}_{-0.2}$ & disc \\
0200980101 & 413.7/397 & $0.17^{+0.03}_{-0.04}$ & $1.5^{+0.3}_{-0.2}$ & $1.5 \pm 0.3$ & $1.3^{+0.4}_{-0.5}$ & $9.4^{+0.5}_{-0.4}$ & disc \\
\\
\multicolumn{8}{c}{Ho IX X-1} \\

0111800101 & 709.8/671 & $0.23 \pm 0.02$ & $1.48^{+0.05}_{-0.06}$ & $2.2 \pm 0.2$ & $1.9 \pm 0.1$ & $24.5 \pm 0.2$ & disc \\
0112521001 & 832.1/819 & $0.15 \pm 0.01$ & $0.31^{+0.04}_{-0.03}$ & $1.65^{+0.03}_{-0.04}$ & $4.1 \pm 0.9$ & $12.4 \pm 0.2$ & HUL \\
0112521101 & 829.8/874 & $0.187 \pm 0.006$ & $<0.01$ & $1.83 \pm 0.02$ & - & $13.8^{+0.4}_{-0.5}$ & HUL \\
0200980101 & 1662.0/1617 & $0.135^{+0.003}_{-0.006}$ & $0.28 \pm 0.01$ & $1.45^{+0.01}_{-0.02}$ & $1.74 \pm 0.09$ & $10.71^{+0.11}_{-0.07}$ & HUL \\
\\
\multicolumn{8}{c}{NGC 4190 ULX1} \\

0654650201 & 488.1/519 & $0.15^{+0.07}_{-0.03}$ & $1.38^{+0.07}_{-0.04}$ & $2.4^{+0.6}_{-0.3}$ & $1.1^{+0.3}_{-0.2}$ & $4.82^{+0.11}_{-0.09}$ & disc \\
0654650301 & 718.6/752 & $0.13 \pm 0.02$ & $1.5^{+0.2}_{-0.1}$ & $1.66^{+0.15}_{-0.09}$ & $2.1 \pm 0.4$ & $8.8 \pm 0.1$ & disc \\
\\
\multicolumn{8}{c}{NGC 4559 ULX2} \\

0152170501 & 439.5/465 & $0.18 \pm 0.02$ & $0.17^{+0.02}_{-0.01}$ & $2.18^{+0.04}_{-0.05}$ & $1.8 \pm 0.3$ & $6.6^{+0.2}_{-0.4}$ & SUL \\
\hline
\end{tabular}
\end{center}
\end{table*}

\begin{table*}
\begin{flushleft} 
{\bf Table \ref{disc_plus_pow}.} (continued)
\end{flushleft}
\begin{center}
\begin{tabular}{cccccccc}
\hline
Obs. ID$^a$ & $\chi^2/{\rm dof} ^b$ &${N_{\rm H}}^c$ & ${kT_{\rm in}}^d$ & $\Gamma^e$ & ${F_{\rm PL}/F_{\rm disc}} ^f$ & ${L_{\rm X}}^g$ & Spectral regime$^h$ \\
 & & & & & (0.3--1 keV) & (0.3--10 keV) & \\
\hline
\multicolumn{8}{c}{NGC 4736 ULX1} \\ 
0404980101 & 541.4/519 & $0.05 \pm 0.02$ & $0.90^{+0.06}_{-0.05}$ & $2.2 \pm 0.1$ & $1.7 \pm 0.3$ & $1.93 \pm 0.03$ & disc \\
\\
\multicolumn{8}{c}{NGC 5204 X-1} \\

0142770101 & 411.5/465 & $0.049 \pm 0.009$ & $0.26 \pm 0.02$ & $1.85 \pm 0.05$ & $2.3 \pm 0.4$ & $4.13^{+0.15}_{-0.09}$ & HUL\\
0405690101 & 665.0/572 & $0.10 \pm 0.01$ & $0.37 \pm 0.03$ & $2.35 \pm 0.08$ & $4 \pm 1$ & $7.8^{+0.2}_{-0.3}$ & SUL \\
0405690201 & 782.0/738 & $0.102^{+0.010}_{-0.009}$ & $0.36^{+0.01}_{-0.02}$ & $2.22^{+0.06}_{-0.07}$ & $2.2 \pm 0.4$ & $7.0 \pm 0.1$ & SUL \\
0405690501 & 626.3/610 & $0.053 \pm 0.008$ & $0.31 \pm 0.02$ & $1.83^{+0.06}_{-0.07}$ & $1.4 \pm 0.2$ & $5.47^{+0.06}_{-0.11}$ & HUL \\
\\
\multicolumn{8}{c}{NGC 5408 X-1} \\

0112290601 & 297.8/317 & $0.02 \pm 0.01$ & $0.22 \pm 0.01$ & $2.5 \pm 0.1$ & $0.6^{+0.2}_{-0.1}$ & $7.6^{+0.4}_{-0.1}$ & SUL \\
0112290701 & 106.1/138 & $0.05 \pm 0.02$ & $0.18 \pm 0.02$ & $2.7 \pm 0.1$ & $1.2^{+0.3}_{-0.2}$ & $8.5 \pm 0.6$ & SUL \\
0302900101 & 1166.3/930 & $0.065^{+0.003}_{-0.004}$ & $0.183^{+0.004}_{-0.003}$ & $2.68^{+0.02}_{-0.03}$ & $1.39^{+0.08}_{-0.05}$ & $6.90^{+0.10}_{-0.07}$ & SUL \\
0500750101 & 763.0/629 & $0.077^{+0.006}_{-0.007}$ & $0.179^{+0.007}_{-0.006}$ & $2.62 \pm 0.04$ & $1.5 \pm 0.1$ & $6.6 \pm 0.2$ & SUL \\
0653380201 & 1448.3/1059 & $0.074 \pm 0.004$ & $0.188^{+0.005}_{-0.004}$ & $2.55 \pm 0.02$ & $1.54^{+0.09}_{-0.05}$ & $7.03^{+0.05}_{-0.03}$ & SUL \\
0653380301 & 1602.6/1126 & $0.068^{+0.003}_{-0.004}$ & $0.194 \pm 0.004$ & $2.56 \pm 0.02$ & $1.66^{+0.10}_{-0.06}$ & $7.79^{+0.04}_{-0.11}$ & SUL \\
0653380401 & 1222.5/1005 & $0.071 \pm 0.004$ & $0.184^{+0.005}_{-0.004}$ & $2.61^{+0.02}_{-0.03}$ & $1.54^{+0.09}_{-0.05}$ & $7.55^{+0.11}_{-0.06}$ & SUL \\
0653380501 & 1246.3/1030 & $0.061^{+0.004}_{-0.003}$ & $0.194^{+0.004}_{-0.005}$ & $2.47^{+0.03}_{-0.02}$ & $1.26^{+0.07}_{-0.04}$ & $7.43^{+0.07}_{-0.12}$ & SUL \\
\\
\multicolumn{8}{c}{NGC 5907 ULX} \\

0145190201 & 152.1/173 & $0.93^{+0.06}_{-0.05}$ & $0.04 \pm 0.02$ & $1.72 \pm 0.05$ & $7 \pm 6$ & $46^{+1}_{-2}$ & HUL \\
0145190101$^j$ & 204.5/168 & $1.5^{+0.2}_{-0.3}$ & $0.13^{+0.02}_{-0.01}$ & $1.87^{+0.08}_{-0.10}$ & $0.4^{+0.3}_{-0.2}$ & $33.3^{+1.8}_{-0.6}$ & HUL \\
0673920301 & 141.5/135 & $0.87^{+0.08}_{-0.06}$ & $0.050^{+0.009}_{-0.007}$ & $1.41^{+0.06}_{-0.05}$ & $1.8 \pm 0.7$ & $16^{+2}_{-1}$ & HUL \\
\\
\multicolumn{8}{c}{NGC 6946 X-1} \\

0200670101 & 35.8/26 & $0.3^{+0.2}_{-0.1}$ & $0.19^{+0.05}_{-0.04}$ & $1.5^{+0.2}_{-0.4}$ & $0.19^{+0.13}_{-0.07}$ & $15^{+10}_{-2}$ & HUL \\
0200670301 & 149.2/157 & $0.12 \pm 0.04$ & $0.21^{+0.03}_{-0.02}$ & $2.3^{+0.1}_{-0.2}$ & $1.0 \pm 0.3$ & $8.6^{+0.6}_{-0.7}$ & SUL \\
0200670401 & 60.3/54 & $<0.1$ & $0.24 \pm 0.05$ & $1.9^{+0.3}_{-0.1}$ & $0.8 \pm 0.3$ & $8.2^{+0.9}_{-0.7}$ & HUL/SUL \\
0500730201 & 425.9/406 & $0.20 \pm 0.03$ & $0.18 \pm 0.01$ & $2.26^{+0.08}_{-0.07}$ & $0.62 \pm 0.07$ & $9.8^{+1.2}_{-0.6}$ & SUL \\
0500730101 & 290.2/273 & $0.07^{+0.03}_{-0.02}$ & $0.23 \pm 0.02$ & $2.05^{+0.09}_{-0.08}$ & $0.8 \pm 0.1$ & $7.9^{+0.2}_{-0.4}$ & HUL/SUL \\

\hline
\end{tabular}
\end{center}
\end{table*}


\begin{table*}
\caption{Fractional variability}
\begin{center}
\begin{tabular}{cccccccc}
\hline
Obs. ID$^a$ & \multicolumn{3}{c}{${F_{\rm var}}^b$} & Obs. ID$^a$ & \multicolumn{3}{c}{${F_{\rm var}}^b$} \\
 & 0.3--10 keV & 0.3--1 keV & 1--10 keV & & 0.3--10 keV & 0.3--1 keV & 1--10 keV \\
\hline
\multicolumn{8}{c}{NGC 55 ULX} \\

0028740201 & $16.2 \pm 0.4$ & $9.1 \pm 0.7$ & $24.0 \pm 0.5$ &
0028740101 & $19.1 \pm 0.7$ & $17 \pm 1$ & $20.6 \pm 0.9$ \\
0655050101 & $8.6 \pm 0.4$ & $8.5 \pm 0.6$ & $8.6 \pm 0.6$ \\
\\
\multicolumn{8}{c}{M31 ULX1} \\

0600660201 & $1.2 \pm 0.7$ & $<2$ & $2.0 \pm 0.7$ &
0600660301 & $<1$ & $<2$ & $<2$ \\
0600660401 & $<2$ & $<3$ & $<3$ &
0600660501 & $<2$ & $<2$ & $<3$ \\
0600660601 & $<5$ & $4 \pm 1$ & $<2$ \\
\\
\multicolumn{8}{c}{NGC 253 XMM2} \\

0125960101 & $24 \pm 2$ & - & $26 \pm 2$ &
0110900101 & $<10$ & - & - \\
0152020101 & $36.6 \pm 0.9$ & $25 \pm 2$ & $42 \pm 1$ &
0304850901 & $11 \pm 4$ & - & $<10$ \\
0304851001 & $7 \pm 4$ & - & $12 \pm 4$ &
0304851201 & $<6$ & - & $<7$ \\
\\
\multicolumn{8}{c}{NGC 253 ULX2} \\

0125960101 & $8 \pm 1$ & - & $9 \pm 1$ &
0110900101 & $5 \pm 3$ & - & $6 \pm 3$ \\
0152020101 & $5 \pm 1$ & - & $5 \pm 2$ &
0304850901 & $7 \pm 3$ & - & $8 \pm 3$ \\
0304851001 & $<9$ & - & $<10$ &
0304851201 & $<10$ & - & $<5$ \\
0304851301 & $11 \pm 4$ & - & $13 \pm 5$ \\
\hline
\end{tabular}
\begin{minipage}{\linewidth}
$^a${\it XMM-Newton} observation identifiers.
$^b$Fractional variability in per cent, calculated from light curves in the appropriate energy band with 200s temporal binning.  Dashes are shown for observations where fractional variability was not calculated in a particular energy band, as there were on average fewer than 20 counts per temporal bin.  Errors and limits correspond to the $1 \sigma$ uncertainty regions.
\end{minipage}
\end{center}
\label{f_var}
\end{table*}

\begin{table*}
\begin{flushleft} 
{\bf Table \ref{f_var}.} (continued)
\end{flushleft}
\begin{center}
\begin{tabular}{cccccccc}
\hline
Obs. ID$^a$ & \multicolumn{3}{c}{${F_{\rm var}}^b$} & Obs. ID$^a$ & \multicolumn{3}{c}{${F_{\rm var}}^b$} \\
 & 0.3--10 keV & 0.3--1 keV & 1--10 keV & & 0.3--10 keV & 0.3--1 keV & 1--10 keV \\
\hline
\multicolumn{8}{c}{M33 X-8} \\

0102640401 & $<3$ & $<9$ & $<3$ &
0102640101 & $1.5 \pm 0.7$ & $<4$ & $<3$ \\
0102640701 & $3 \pm 2$ & $5 \pm 3$ & $3 \pm 2$ &
0102641001 & $3 \pm 2$ & $9 \pm 3$ & $<5$ \\
0102642001 & $<2$ & $<6$ & $<3$ &
0102642101 & $1.6 \pm 0.9$ & $<7$ & $2 \pm 1$ \\
0102642301 & $1.6 \pm 0.8$ & $<3$ & $<3$ &
0141980601 & $<2$ & $<4$ & $<2$ \\
0141980801 & $<1$ & $<2$ & $<2$ &
0141980101 & $<4$ & $7 \pm 3$ & $<5$ \\
0141980301 & $<2$ & $<5$ & $<4$ &
0650510101 & $3.1 \pm 0.3$ & $4.5 \pm 0.7$ & $2.4 \pm 0.5$ \\
0650510201 & $0.7 \pm 0.5$ & $1.3 \pm 0.8$ & $1.3 \pm 0.4$ \\
\\
\multicolumn{8}{c}{NGC 1313 X-1} \\

0106860101 & $6 \pm 1$ & $6 \pm 3$ & $7 \pm 2$ &
0150280301 & $5 \pm 2$ & $12 \pm 3$ & $<5$ \\
0150280601 & $<5$ & $<6$ & $4 \pm 2$ &
0205230301 & $5 \pm 1$ & $5 \pm 3$ & $6 \pm 1$ \\
0205230501 & $<7$ & - & $<9$ &
0205230601 & $<4$ & $<12$ & $<5$ \\
0405090101 & $2.6 \pm 0.8$ & $<3$ & $3.4 \pm 0.9$ \\
\\
\multicolumn{8}{c}{NGC 1313 X-2} \\

0106860101 & $<5$ & $<9$ & $5 \pm 4$ &
0150280301 & $8 \pm 1$ & $<7$ & $8 \pm 1$ \\
0150280601 & $4 \pm 3$ & $<8$ & $7 \pm 3$ &
0205230301 & $<3$ & $<6$ & $<5$ \\
0205230501 & $<7$ & $<8$ & $<5$ &
0205230601 & $5 \pm 2$ & $6 \pm 5$ & $4 \pm 2$ \\
0301860101 & $<3$ & $7 \pm 3$ & $<3$ &
0405090101 & $9.1 \pm 0.4$ & $<8$ & $10.7 \pm 0.5$ \\
\\
\multicolumn{8}{c}{IC 342 X-1} \\

0093640901 & $5 \pm 3$ & - & $6 \pm 3$ &
0206890201 & $<6$ & - & $3 \pm 2$ \\
\\
\multicolumn{8}{c}{IC 342 X-2} \\

0093640901 & $<8$ & - & $<8$ &
0206890201 & $5 \pm 2$ & - & $5 \pm 2$ \\
\\
\multicolumn{8}{c}{NGC 2403 X-1} \\

0164560901 & $<6$ & - & $<10$ \\
\\
\multicolumn{8}{c}{Ho II X-1} \\

0112520601 & $<3$ & $3 \pm 1$ & $<3$ &
0200470101 & $2.0 \pm 0.5$ & $<2$ & $3.1 \pm 0.7$ \\
0561580401 & $20.7 \pm 0.5$ & $20.0 \pm 0.7$ & $21.2 \pm 0.8$ \\
\\
\multicolumn{8}{c}{M81 X-6} \\

0111800101 & $6.1 \pm 0.7$ & $8 \pm 2$ & $6.5 \pm 0.9$ &
0112521001 & $<5$ & - & $<6$ \\
0112521101 & $7 \pm 2$ & $11 \pm 7$ & $<8$ &
0200980101 & $<7$ & - & $<7$ \\
\\
\multicolumn{8}{c}{Ho IX X-1} \\

0111800101 & $<2$ & $5 \pm 3$ & $<4$ &
0112521001 & $<2$ & $<4$ & $<2$ \\
0112521101 & $<2$ & $<3$ & $<4$ &
0200980101 & $<2$ & $<3$ & $2 \pm 1$ \\
\\
\multicolumn{8}{c}{NGC 4190 ULX1} \\

0654650201 & $<3$ & $<9$ & $<4$ &
0654650301 & $<2$ & $<4$ & $<3$ \\
\\
\multicolumn{8}{c}{NGC 4559 ULX2} \\

0152170501 & $7 \pm 2$ & $9 \pm 2$ & $6 \pm 3$ \\
\\
\multicolumn{8}{c}{NGC 4736 ULX1} \\

0404980101 & $16 \pm 1$ & $6 \pm 3$ & $22 \pm 1$ \\
\\
\multicolumn{8}{c}{NGC 5204 X-1} \\

0142770101 & $4 \pm 2$ & $<5$ & $4 \pm 3$ &
0405690101 & $<4$ & $5 \pm 2$ & $3 \pm 2$ \\
0405690201 & $3 \pm 1$ & $3 \pm 2$ & $<10$ &
0405690501 & $<3$ & $<4$ & $<6$ \\
\\
\multicolumn{8}{c}{NGC 5408 X-1} \\

0112290601 & $5 \pm 1$ & $3 \pm 2$ & $7 \pm 3$ &
0112290701 & $<4$ & $<9$ & $<7$ \\
0302900101 & $9.8 \pm 0.3$ & $7.2 \pm 0.5$ & $15.4 \pm 0.6$ &
0500750101 & $14.9 \pm 0.6$ & $10.4 \pm 0.8$ & $23 \pm 1$ \\
0653380201 & $6.3 \pm 0.4$ & $4.5 \pm 0.6$ & $9.4 \pm 0.7$ &
0653380301 & $7.1 \pm 0.3$ & $5.5 \pm 0.5$ & $10.2 \pm 0.5$ \\
0653380401 & $8.6 \pm 0.3$ & $6.0 \pm 0.5$ & $13.2 \pm 0.6$ &
0653380501 & $8.2 \pm 0.3$ & $4.6 \pm 0.6$ & $13.4 \pm 0.6$ \\
\hline
\end{tabular}
\end{center}
\end{table*}

\begin{table*}
\begin{flushleft} 
{\bf Table \ref{f_var}.} (continued)
\end{flushleft}
\begin{center}
\begin{tabular}{cccccccc}
\hline
Obs. ID$^a$ & \multicolumn{3}{c}{${F_{\rm var}}^b$} & Obs. ID$^a$ & \multicolumn{3}{c}{${F_{\rm var}}^b$} \\
 & 0.3--10 keV & 0.3--1 keV & 1--10 keV & & 0.3--10 keV & 0.3--1 keV & 1--10 keV \\
\hline
\multicolumn{8}{c}{NGC 5907 ULX} \\

0145190201 & $8 \pm 2$ & - & $8 \pm 2$ &
0145190101 & $10 \pm 2$ & - & $10 \pm 2$ \\
0673920301 & $9 \pm 4$ & - & $7 \pm 5$ \\
\\
\multicolumn{8}{c}{NGC 6946 X-1} \\

0200670101 & $10 \pm 6$ & $<20$ & $<10$ &
0200670301 & $14 \pm 2$ & $<8$ & $18 \pm 3$ \\
0200670401 & $<8$ & - & $12 \pm 6$ &
0500730201 & $24 \pm 1$ & $13 \pm 2$ & $33 \pm 1$ \\
0500730101 & $27 \pm 1$ & $15 \pm 2$ & $35 \pm 2$ \\
\hline
\end{tabular}
\end{center}
\end{table*}

\bsp

\label{lastpage}

\end{document}